\documentclass[english]{revtex4-1}
\usepackage[T1]{fontenc}
\usepackage[latin9]{inputenc}
\usepackage{color}
\definecolor{page_backgroundcolor}{rgb}{1, 1, 1}
\pagecolor{page_backgroundcolor}
\usepackage{float}
\usepackage{textcomp}
\usepackage{amsmath}
\usepackage{amssymb}
\usepackage{graphicx}

\usepackage{makeidx}

\makeatletter

\@ifundefined{textcolor}{}
{%
 \definecolor{BLACK}{gray}{0}
 \definecolor{WHITE}{gray}{1}
 \definecolor{RED}{rgb}{1,0,0}
 \definecolor{GREEN}{rgb}{0,1,0}
 \definecolor{BLUE}{rgb}{0,0,1}
 \definecolor{CYAN}{cmyk}{1,0,0,0}
 \definecolor{MAGENTA}{cmyk}{0,1,0,0}
 \definecolor{YELLOW}{cmyk}{0,0,1,0}
}

\usepackage{babel}

\usepackage{babel}

\usepackage{indentfirst}

\makeatother

\usepackage{babel}
\begin{document}

\title{EFFECTS OF DYNAMIC SYNAPSES ON NOISE-DELAYED RESPONSE LATENCY OF
A SINGLE NEURON }

\author{M. Uzuntarla$^{\dagger,}$$^{\ast}$
, M. Ozer$^{\ddagger}$, U. Ileri$^{\dagger}$, A.Calim$^{\dagger}$,
and J. J. Torres$^{\flat}${\small{}}\\
{\small{}$\dagger$ Department of Biomedical Engineering, Bulent Ecevit
University, }\\
{\small{}Engineering Faculty, 67100 Zonguldak, Turkey}\\
{\small{}$\ddagger$ Department of Electrical and Electronics Engineering,
Bulent Ecevit University, }\\
{\small{}Engineering Faculty, 67100 Zonguldak, Turkey}\\
{\small{}$\flat$ Department of Electromagnetism and Physics of the
Matter and }\\
{\small{}Institute Carlos I for Theoretical and Computational Physics,}\\
{\small{} University of Granada, Granada, E-18071 Spain}\\
{\small{}$\ast$ The Krasnow Institute for Advanced Study, George
Mason University, }\\
{\small{}Fairfax, Virginia 22030, USA}}
\begin{abstract}
Noise-delayed decay (NDD) phenomenon emerges when the first-spike
latency of a periodically forced stochastic neuron exhibits a maximum
for a particular range of noise intensity. Here, we investigate the
latency response dynamics of a single Hodgkin-Huxley neuron that is
subject to both a suprathreshold periodic stimulus and a background
activity arriving through dynamic synapses. We study the first spike
latency response as a function of the presynaptic firing rate $f$.
This constitutes a more realistic scenario than previous works, since
$f$ provides a suitable biophysically realistic parameter to control
the level of activity in actual neural systems. We first report on
the emergence of classical NDD behavior as a function of $f$ for
the limit of static synapses. Secondly, we show that when short-term
depression and facilitation mechanisms are included at synapses,
different NDD features can be found due to the their modulatory effect
on synaptic current fluctuations. For example a new intriguing double
NDD (DNDD) behavior occurs for different sets of relevant synaptic
parameters. Moreover, depending on the balance between synaptic depression
and synaptic facilitation, single NDD or DNDD can prevails, in such
a way that synaptic facilitation favors the emergence of DNDD whereas
synaptic depression favors the existence of single NDD. This is the
first time it has been reported the existence of DNDD effect in response
latency dynamics of a neuron.\\ 

\noindent PACS number(s): 87.19.lc, 05.40.\textminus a, 05.45.\textminus a

\end{abstract}
\maketitle

\section{Introduction}

Despite the efforts made over the last decades to clarify the functions
of actual neural systems, one of the most important questions not
yet addressed is to understand how neural coding naturally occurs.
Although the coding mechanism used by the neurons is still unclear,
it is widely assumed that information coding is based on action potentials
(APs) or \emph{spikes.} In the context of spike based communication,
a potential coding mechanism is that, under strong temporal constraints,
neurons may perform information processing with only one spike using
first-spike latency as an information carrier \cite{Eckmiller:90,vanrullen_spike_2005}.
First-spike latency coding has been found as a meaningful strategy
because it provides a fast as well as an energy efficient environment
for information processing in actual neural systems. In fact, numerous
experimental works conducted in peripheral and central neurons have
demonstrated that first spike latency duration carries a greater amount
of information about the received stimulus features than subsequent
spikes \cite{chase2007first,panzeri2001role,junek2010olfactory,furukawa2002cortical,heil2004first,gawne1996latency,reich2001temporal,vanrullen_spike_2005}. 

Beside these experimental findings, theoretical and computational
studies have also been performed to investigate the influence of different
biophysical mechanisms shaping first-spike latency response of neurons.
In this context, Pankratova et al. \cite{pankratova2005a,pankratova2005b}
theoretically investigated the impact of noise -- which is ubiquitous
in whole nervous system -- on latency dynamics of a single neuron
that is subject to a suprathreshold periodic input current. They reported
that for a range of periodic forcing frequency near the boundaries
of suprathreshold spiking regime, there exists a resonance-like behavior
of the mean latency depending on noise intensity. More precisely,
as noise increases, mean response latency first dramatically increases,
then reaches some maximum, and finally decreases to a value lower
than the one in noise-free condition. The authors called this non-monotonic
noise dependence of mean latency \textquotedblleft Noise Delayed Decay
(NDD)\textquotedblright . This phenomenon indicates that latency coding
is not a convenient strategy to encode signals near the spiking threshold
because it implies, first, a delay in external signal detection and,
second, a very low temporal spiking precision for a particular intensity
of noise. In \cite{ozer2008,Ozer2008b}, NDD was also investigated
on the level of network, and it was reported that NDD effect on latency
response can be controlled via network activity. However, these studies
\cite{pankratova2005a,pankratova2005b,ozer2008,Ozer2008b} do not
motivate, from a biophysical point of view, the source of the noise
and it was incorporated artificially by adding an external additive
stochastic input current. Therefore, these works can not give a biophysical
explanation for the emergence of NDD. Recently, more realistic assumptions
have been considered to characterize the noise, including ion channel
noise \cite{ozer2009}-- which allows to relate intrinsic dynamics
of a neuron with NDD -- and synaptic background activity -- that allows
to relate unreliability of spike transmission with NDD \cite{uzuntarla2012}. 

In the present study, we investigate NDD phenomenon in a more realistic
scenario considering \textit{dynamic} synapses. In fact, modeling
synapses as static connections (with fixed conductance) does not reveal
the possible influence of some biophysical processes on NDD. It is
well-known, for instance, that synapses exhibit a high variability
with a diverse origin during information transmission, such as stochastic
release of neurotransmitters, variations in chemical concentration
through synapses and spatial heterogeneity of synaptic response over
dendrite tree \cite{fssjn03}. The collective effect of all these
factors might result in synaptic conductance fluctuations at short
time scales. It is known, e.g., that synapses in different cortical
areas can have varied forms of plasticity, being either in only a
specific form, or showing a mixture of several forms \cite{tsodyksCODING,markramPNAS,markram06}.
For instance, postsynaptic response can be depressed or facilitated
during high presynaptic activity. Synaptic depression is induced due
to the fact that available neurotransmitter concentration at synaptic
buttons needs some time to recover after each release event originated
by arrival of presynaptic APs. If APs arrive at high frequency, the
release probability of neurotransmitters for subsequent APs will decrease,
and therefore postsynaptic response will be decreased or depressed.
On the other hand, synaptic facilitation is a consequence of residual
cytosolic calcium -- that remains inside the synaptic buttons after
the arrival of the first APs -- which favors the release of more neurotransmitter
vesicles for the next arriving AP \cite{bertramJNEURO}. Such increase
in neurotransmitters causes a potentiation of the postsynaptic response
or synaptic facilitation. Short-term depression (STD) and short-term
facilitation (STF) mechanisms have been reported to be relevant for
various brain functions, i.e. cortical gain control \cite{abott1997},
information storage in neural networks \cite{bibitchkov2002,torres2002,mejias09},
coincidence detection of signals \cite{pantic2003,mejiasCD08}, synchrony
and selective attention \cite{Sejnowski2006,BT05}, and perform new
computations in diverse neural systems \cite{TorresKappen2013}.

Our results in the present work reveal that dynamic synapses with
STD and STF mechanisms might give rise to emergence of rich NDD features
in response latency of a single neuron. We find, e.g., pure depressing
synapses, with low levels of depression, induce a new intriguing double
NDD (DNDD) behavior which collapses into a single NDD as the level
of depression increases. This is the first time it has been reported
the existence of DNDD behavior as a function of a biophysically realistic
parameter controlling the level of activity in a neural medium.
We also show that NDD effect can be attenuated or even completely
eliminated from the latency response of a neuron when the level of
depression is very high. On the other hand, in the presence of STF
mechanism competing with STD and depending on the balance between
these two mechanisms, both single NDD or DNDD can emerge, in such
a way that synaptic facilitation favors the emergence of DNDD whereas
synaptic depression favors the existence of single NDD. We finally
clarify the underlying mechanism that gives rise to NDD and DNDD behavior
in terms of the non-monotonic dependence of synaptic current fluctuations
on presynaptic activity due to the presence of dynamic synapses.

\section{Models and Methods}

A schematic illustration of the system under study is depicted in
Fig. 1. This consists of a postsynaptic neuron which is driven by
a suprathreshold periodic stimulus and subject to an uncorrelated
presynaptic activity from a finite number of excitatory and inhibitory
neurons.

\begin{figure}[H]
\begin{centering}
\includegraphics[scale=0.5]{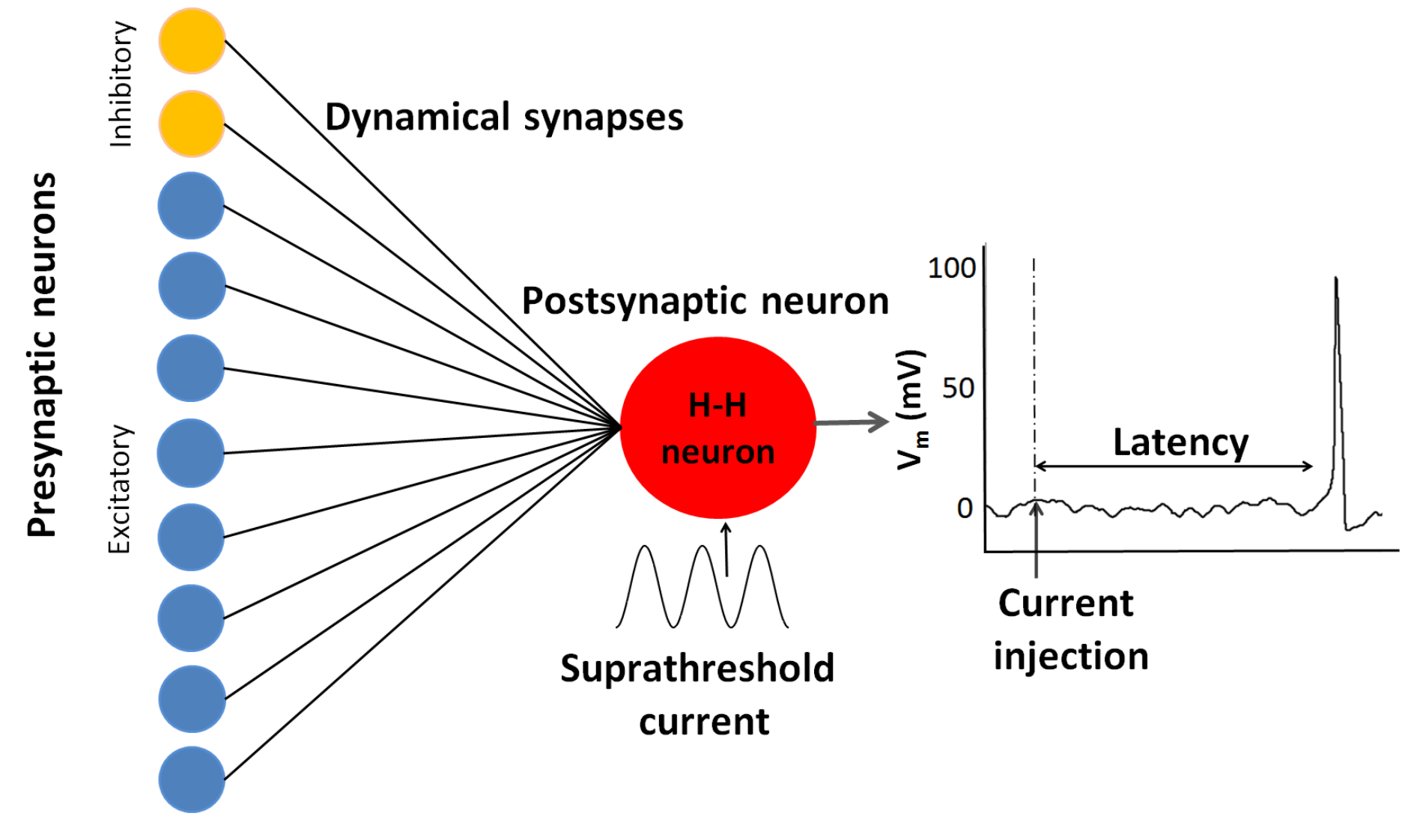}
\par\end{centering}

\protect\caption{(Color online) Schematic illustration of the considered system. The
postsynaptic neuron is subject to a suprathreshold periodic current
and a noisy background activity resulting from random arrival of excitatory
and inhibitory presynaptic spikes. Our aim here is to investigate
how the dynamic properties of these synapses can influence latency
response of the postsynaptic neuron.}
\end{figure}
The time evolution of membrane potential of the postsynaptic neuron
is modeled based on the Hodgkin-Huxley (H-H) equations as follows
\cite{HHb}:

\noindent 
\begin{equation}
C_{m}\frac{dV_{m}}{dt}=-G_{Na}m^{3}h\left(V_{m}-E_{Na}\right)-G_{K}n^{4}\left(V_{m}-E_{K}\right)-G_{L}\left(V_{m}-E_{L}\right)+I_{stim}+I_{syn}
\end{equation}

\noindent where $V_{m}$ is the membrane potential, $C_{m}=1\,\mu F/cm^{2}$
is the membrane capacity per unit area. The constants $G_{Na}=120\, mS/cm^{2}$,
$G_{K}=36\, mS/cm^{2}$ and $G_{L}=0.3\, mS/cm^{2}$ are maximum conductance
for sodium, potassium and leakage channels, respectively. $E_{Na}=115\, mV$,
$E_{K}=-12\, mV$ and $E_{L}=10.6\, mV$ denote the corresponding
sodium, potassium and leak current reversal potentials. The gating
variables $m$, $h$ and $n$ that govern activation and inactivation
of sodium channels and activation of potassium channels, respectively,
and obey the following differential equations \cite{HHb}:
\noindent 
\begin{equation}
\frac{d\gamma}{dt}=\alpha_{\gamma}(V_{m})(1-\gamma)-\beta_{\gamma}(V_{m})\gamma
\end{equation}
\noindent where $\alpha_{\gamma}$ and $\beta_{\gamma}$ $(\gamma=m,\, n,\, h)$
are experimentally determined voltage dependent rate functions for
the gating variable $\gamma$ and can be found in \cite{pankratova2005b}.
In Eq. (1), $I_{stim}$ is the suprathreshold periodic current stimulus,
$I_{stim}(t)=A_{0}sin(2\pi f_{s}t)$ where $A_{0}$ and $f_{s}$ stand,
respectively, for the amplitude and frequency of this signal. Here
``suprathreshold'' refers to an input level that produces tonic
spiking activity. Finally, $I_{syn}$ represents the total synaptic
current generated by $N=1000$ presynaptic excitatory and inhibitory
neurons where excitatory to inhibitory ratio is taken as $N_{e}:N_{i}=4:1$,
a choice that preserves the similar ratio found in the mammalian cortex
\cite{Braitenberg91}. Each presynaptic neuron is considered as an
independent Poisson spike train generator with the same presynaptic
firing rate $f$.

Synaptic connections between presynaptic and postsynaptic neurons
are modeled according to Tsodyks and Markram dynamical synapse equations
\cite{tsodyksNC}. Although short-term synaptic plasticity is widely
considered to be valid for excitatory synapses, recent experimental
studies have shown that inhibitory synapses also display this type
of plasticity behavior \cite{Fitzpatrick2001,Tecuapetla2007,Ma2012,Flores2015}.
Thus, in our study, it is reasonable to assume that Tsodyks and Markram
equations can also be used to model inhibitory synapses, as previously
considered in \cite{tsodyksNC,tsodyksjn00}. The model assumes that
an AP can be transmitted by a finite amount of neurotransmitter resources.
That is, each presynaptic AP at a given synapse $i$ activates a fraction
of neurotransmitters with probability $u_{i}(t)$ (release probability),
which then quickly inactivates (as a fast synapse) during a characteristic
time constant $\tau_{in}=3\, ms$ \cite{tsodyksPNAS,tsodyksNC}. After
a recovery period $\tau_{\ensuremath{rec}}$, resources are reloaded
and the synapse $i$ returns to its initial state. This process is
governed by the following equations \cite{tsodyksNC}:

\noindent 
\begin{equation}
\frac{dx_{i}(t)}{dt}=\frac{z_{i}(t)}{\tau_{rec}}-u_{i}(t)x_{i}(t)\delta(t-t_{spk}^{i})\;\;
\end{equation}
\begin{equation}
\frac{dy_{i}(t)}{dt}=-\frac{y_{i}(t)}{\tau_{in}}+u_{i}(t)x_{i}(t)\delta(t-t_{spk}^{i})
\end{equation}
\begin{equation}
\frac{dz_{i}(t)}{dt}=\frac{y_{i}(t)}{\tau_{in}}-\frac{z_{i}(t)}{\tau_{rec}}\quad\quad\quad\quad\quad\quad\quad
\end{equation}
\\
\noindent where $x_{i},\, y_{i},\, z_{i}$ are the fraction of neurotransmitters
in a recovered, active and inactive state, respectively. Delta functions
in equations take into account that an AP arrives to the synapse at
$t=t_{spk}^{i}$. The model described by Eqs. $(3-5)$ satisfactorily
explains STD mechanism in cortical neurons for relatively large values
of $\tau_{rec}$ and assuming $u_{i}(t)=\mathcal{U}$, which represents
a constant neurotransmitter release probability after arrival of an
AP \cite{tsodyksPNAS,tsodyksNC}. On the other hand, STF mechanism
can be introduced by assuming that $u_{i}(t)$ is not fixed but is
increased by a certain amount due to the influx of calcium ions through
voltage sensitive ion channels into the presynaptic neuron when an
AP arrives. The corresponding dynamic equation for $u_{i}(t)$ is
given by \cite{tsodyksNC}:

\noindent 
\begin{equation}
\frac{du_{i}(t)}{dt}=\frac{\mathcal{U}-u_{i}(t)}{\tau_{fac}}+\mathcal{U}[1-u_{i}(t)]\delta(t-t_{spk}^{i})
\end{equation}

\noindent where $\tau_{\ensuremath{fac}}$ is the time duration of transition
from open to close state of calcium channel gates \cite{tsodyksNC}.
It is worth noting that large values of $\tau_{\ensuremath{rec}}$
and $\tau_{\ensuremath{fac}}$ are associated to stronger synaptic
depression and facilitation at the synapse, respectively. More precisely,
large $\tau_{rec}$ implies that neurotransmitter concentration $x(t)$
in ready-releasable pool takes more time to recover its maximum value
after the release event induced by an AP. Thus, if another AP arrives
at certain posterior time $t_{spk}^{i}$, the amount of released neurotransmitters,
that is $u_{i}(t_{spk}^{i})x_{i}(t_{spk}^{i})$, will be lower and
therefore the postsynaptic response for the second AP too. This depressing
effect will be stronger for larger $\tau_{rec}$ and higher presynaptic
firing frequency. On the other hand, large $\tau_{fac}$ in Eq. (6)
implies that, assuming an initial low value of the release probability
$u_{i}(t),$ it will take a large time to grow until the value $\mathcal{U}$
indicating that the facilitation effect (increase of the release probability)
will take place during a large period of time. Finally, for a given
value of $\tau_{rec},$ a large value of $\mathcal{U}$ induces a
strong depletion of the available resources for a given AP, and therefore
a decrease of the postsynaptic response for subsequent APs, particularly
at high presynaptic frequency. In the case of facilitating synapses,
for a given value of $\tau_{fac},$ a larger value for $\mathcal{U}$
produces a stronger increase in the release probability after each
AP so it induces a strong and fast facilitation for high-frequency
presynaptic activity. Within this model, the postsynaptic current
generated at a synapse $i$ is taken to be proportional to the fraction
of neurotransmitters in the active state, namely $I_{i}(t)=\mathcal{A}y_{i}(t).$
Here, $\mathcal{A}$ is the maximum synaptic current which can be
generated at the synapse only by activating all resources. 

Based on the description of the dynamic synapse model on the level
of single synapse in Eqs. $(3-6)$, we now can generalize the total
synaptic current generated by excitatory and inhibitory presynaptic
neurons as follows:

\noindent 
\begin{equation}
I_{syn}(t)=\sum_{p=1}^{N_{E}}\mathcal{A}y_{p}(t)-K\sum_{q=1}^{N_{i}}\mathcal{A}y_{q}(t)
\end{equation}
\noindent where $K$ is the relative strength between inhibitory and excitatory
connections, and $-K\mathcal{A}$ as a whole stands for maximum inhibitory
synaptic current per synapse. Here, we assume $K=4$ within the physiological
range of balanced state of cortical neurons \cite{Braitenberg91}. 

To evaluate the emergence of NDD phenomenon in response latency dynamics
of the postsynaptic neuron, we define latency to the first spike as
the time when an AP first crosses with a positive slope a detection
threshold, taken here equal to $20\, mV$, relative to start of the
stimulus (see Fig. 1). Then, we compute mean latency of an ensemble
of first spikes by averaging over $r$ realizations:

\noindent 
\begin{equation}
Mean\, Latency=\langle t\rangle=\frac{1}{r}\sum_{i=1}^{r}t_{i}
\end{equation}

\noindent where $t_{i}$ is the appearance time of the first spike for $i$th
realization. We also consider standard deviations of latencies, or
temporal jitter, as follows:

\noindent 
\begin{equation}
Jitter=\sqrt{\langle t^{2}\rangle-\langle t\rangle^{2}}
\end{equation}

\noindent where $\langle t^{2}\rangle$ represents the mean squared
latency. The results presented in the next sections are obtained over
$r=5000$ independent runs for each set of parameter values to warrant
appropriate statistical accuracy with respect to the stochastic fluctuations
in background activity. Numerical integration of the whole system
equations are performed using standard fourth order Runge-Kutta algorithm
with a step size of $10\,\mu s$.

\section{Results}

\subsection{Emergence of NDD behavior with dynamic synapses}

In this section, we have systematically analyzed the emergence of
NDD in a H-H neuron which is subject to both a suprathreshold periodic
signal and a noisy synaptic background activity arising from a balanced
population of excitatory and inhibitory neurons. In previous works
on NDD \cite{pankratova2005a,pankratova2005b,ozer2009}, it has been
illustrated that for a given value of periodic signal amplitude $A_{0}$,
a H-H neuron operates in a suprathreshold regime only for a range
of signal frequencies $f_{s}$. In these works, it has also been demonstrated
that NDD effect only emerges near the lower and upper limits of such
frequency regions. As an example, we consider here the case of $A_{0}=4\,\mu A/cm^{2}$
that implies a frequency range $f_{s}\in[16:149]\, Hz$ for suprathreshold
regime (see Fig. 1 in \cite{pankratova2005b}). To investigate NDD
phenomenon, we also fix the stimulus frequency to $f_{s}=20\, Hz$
which is just above the lower limit of the suprathreshold frequency
region. Here, it is worth noting that similar qualitative results
can be obtained if one considers a signal frequency near the upper
limit of considered $f_{s}$ region. 

First, we have investigated the influence of short-term synaptic depression
(varying $\tau_{rec}$) on the emergence of NDD phenomenon by measuring
latency response statistics of the postsynaptic neuron as a function
of presynaptic firing rate $f$, which is considered here as global
control parameter to scale the intensity of background activity. Since
we assume that postsynaptic neuron receives inputs through purely
depressing synapses, we set $\tau_{fac}=0$ to block the facilitation
mechanism in the model. In Fig. 2, mean latency and jitter are plotted
versus $f$ for several values of $\tau_{rec}$ including the case
of \emph{static} or non-depressed synapses $(\tau_{rec}=0).$ Our
results show that mean latency and jitter exhibit a non-monotonic
behavior, i. e. NDD phenomenon, as a function of $f$, both for static
and dynamic synapses with relatively large values of $\tau_{rec}$
(see Panels A and B in Fig. 2). 

\begin{figure}[H]
\begin{centering}
\includegraphics[scale=0.25]{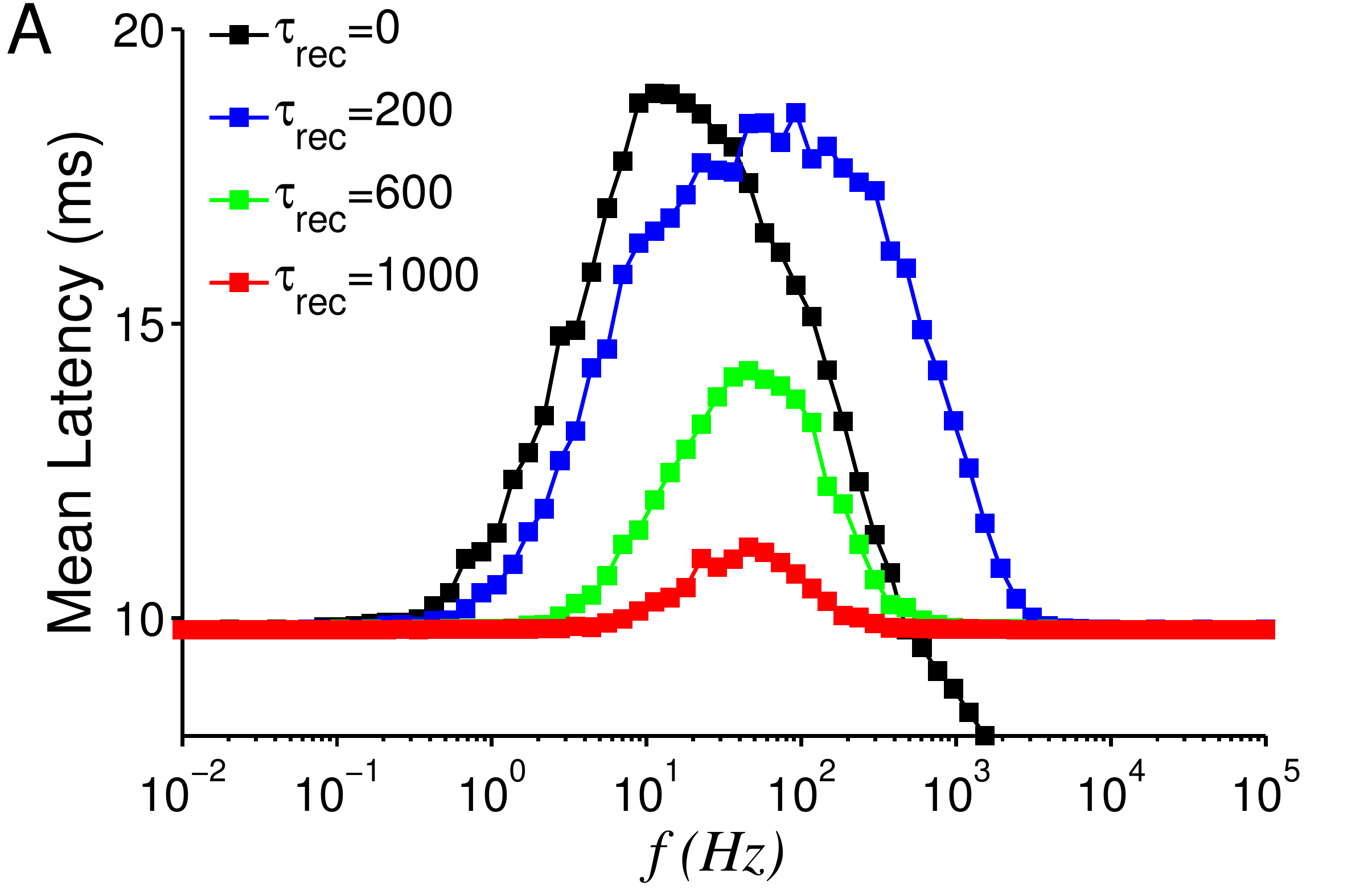}\includegraphics[scale=0.25]{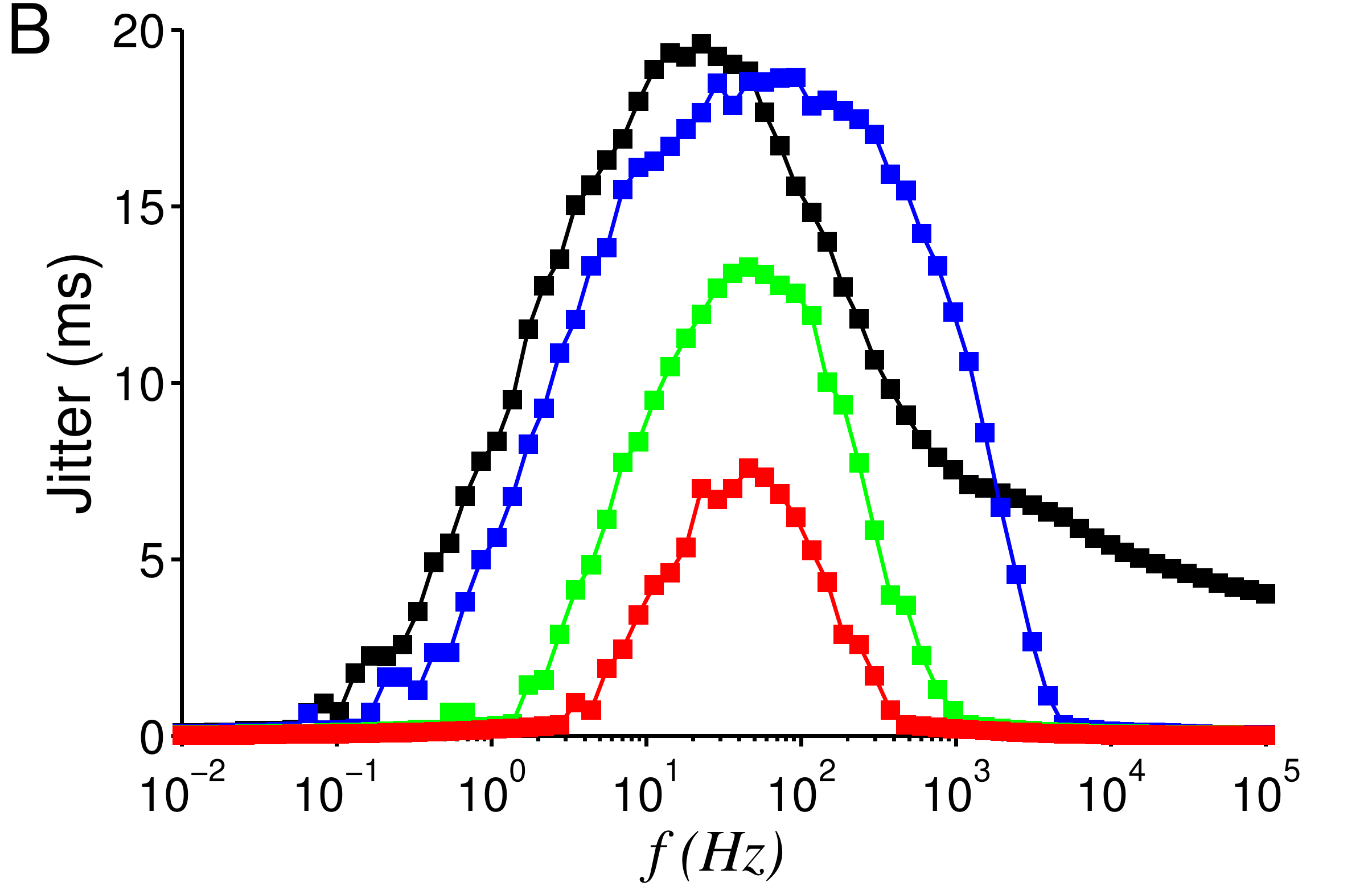}\vspace{1mm}
\par\end{centering}
\begin{centering}
\includegraphics[scale=0.25]{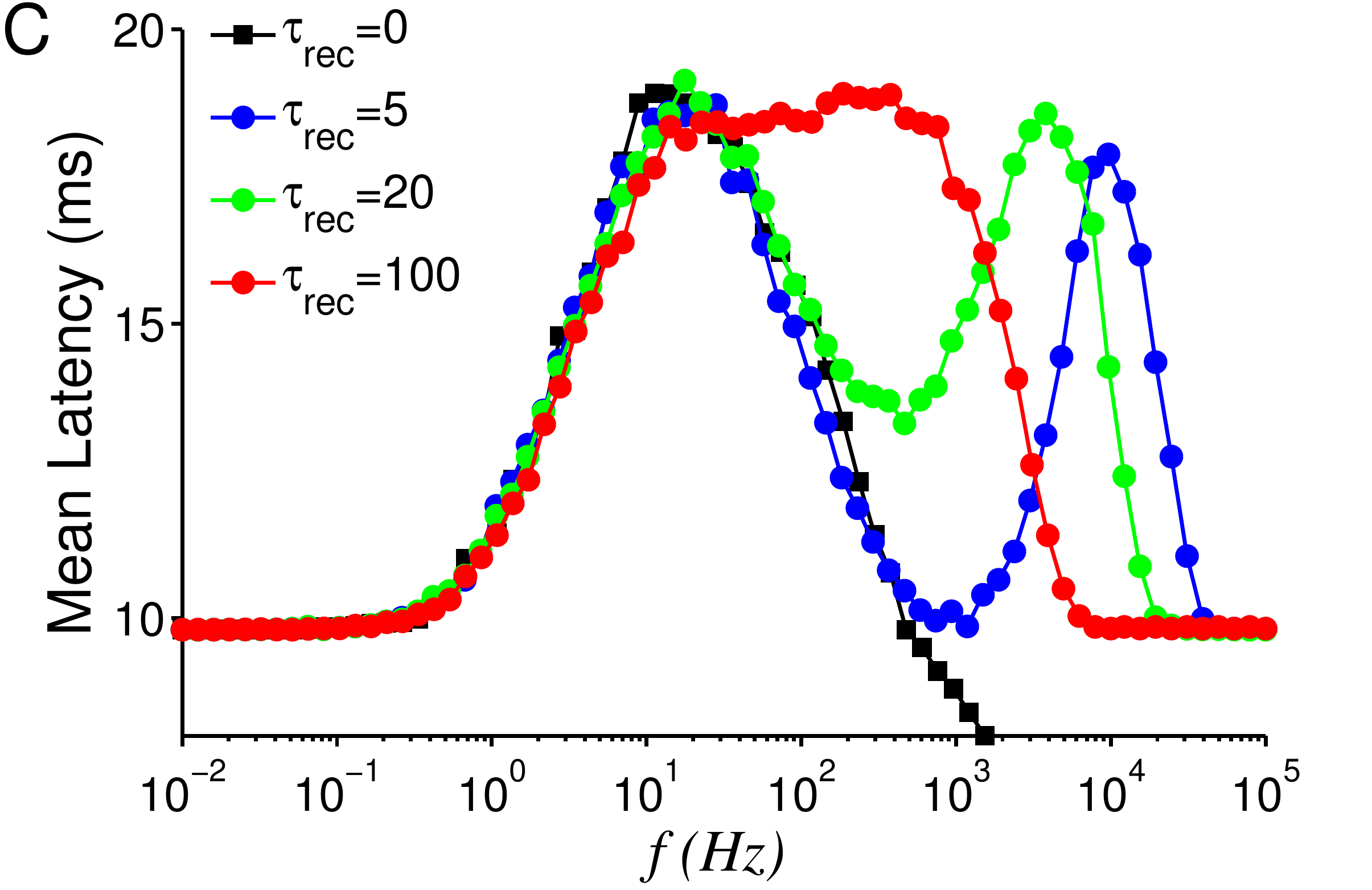}\includegraphics[scale=0.25]{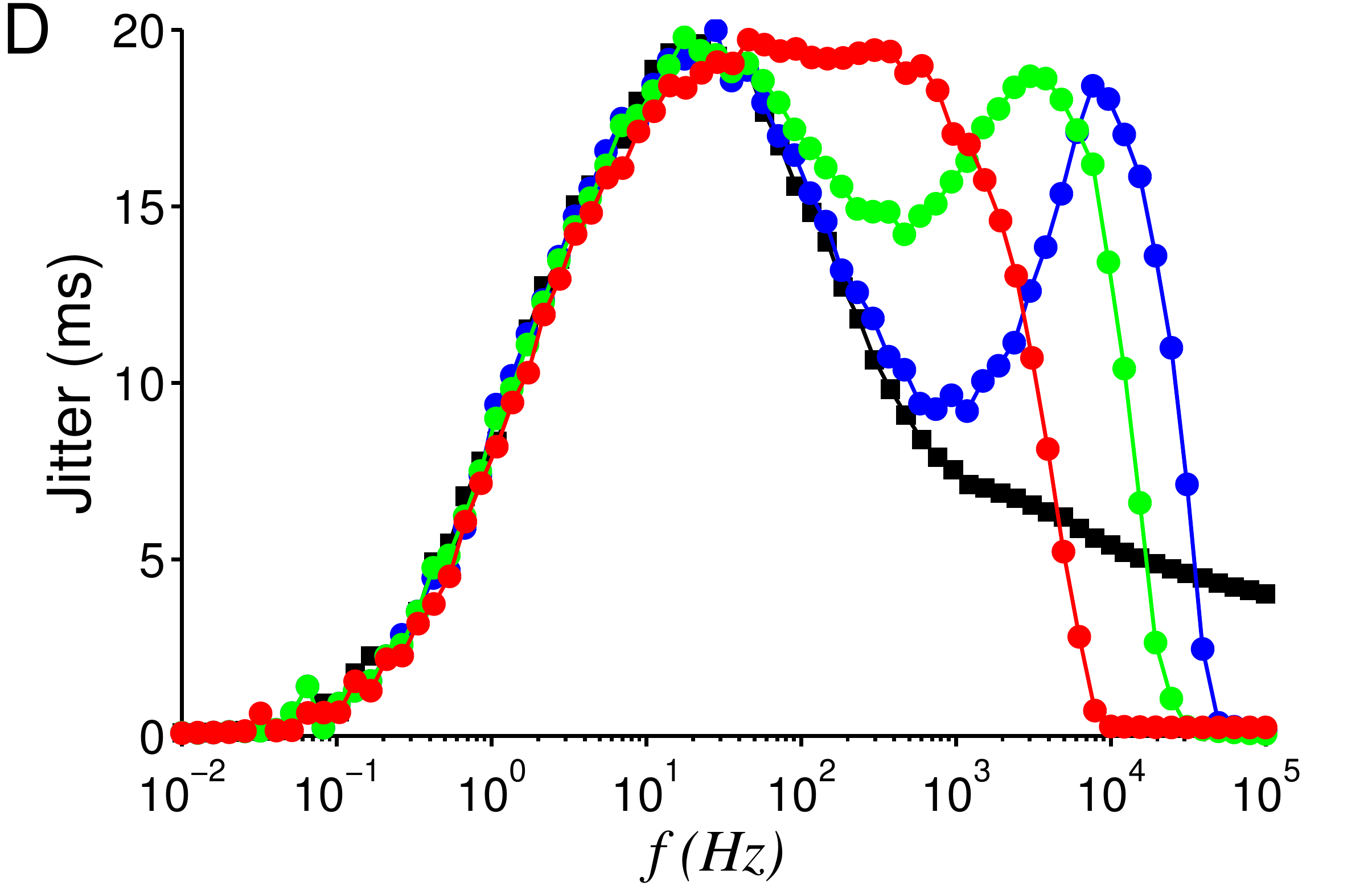}
\par\end{centering}
\centering{}\protect\caption{(Color online) The influence of short term synaptic depression on
first-spike latency response statistics of a H-H neuron. Figure depicts
the behavior of mean latency (left panels) and jitter (right panels)
against presynaptic firing frequency $f$ for several values of $\tau_{rec}$.
NDD behavior emerges for depressed $(\tau_{rec}>0)$ and non-depressed
$(\tau_{rec}=0)$ synapses. This unwanted effect appearing for a wide
range of $f$ can disappear for relatively large levels of depression
as seen in panels A and B. On the other hand, panels C and D illustrate,
respectively, the dependence of statistics for small values of $\tau_{rec}$.
When the depression is lowered, interestingly, a double NDD behavior
emerges. Other synaptic parameters were $\mathcal{A}=0.6\, nA$, $\mathcal{\mathcal{U}}=0.1$,
$\tau_{fac}=0$. }
\end{figure}

In addition to the classical NDD, Fig. 2 also depicts intriguing effect
of synaptic depression on NDD curves. That is how synaptic depression
can modulate the shape, amplitude and width of the NDD curves. For
instance, a double resonance like regime seems to emerge for the mean
latency and jitter in considered $f$ domain for relatively low levels
of synaptic depression $\tau_{rec}<100\, ms$ (see Panels C and D
in Fig. 2). To the best of our knowledge, this is the first time such
intriguing ``double NDD'' (DNDD) effect has been described. DNDD
behavior is more evident for very low $\tau_{rec}$ and results in
a single NDD curve when $\tau_{rec}$ approaches to a value of $100\, ms$.
On the other hand, for $\tau_{rec}\thickapprox100\, ms$, the resulting
merged single NDD curve presents the maximum possible amplitude, as
in the classical NDD behavior for static synapses, but with a very
large width in $f$ domain compared with the static synapse case.
This implies, therefore, a poor first spike latency coding efficiency
for this range of depression. Finally, for $\tau_{rec}>100\, ms$,
mean latency and jitter start to decrease in amplitude and width as
the level of depression increases, which might provide a relevant
mechanism to increase the first spike latency coding efficiency by
increasing the level of depression at synapses. In others words, since
the existence of NDD implies a delay in signal detection by a given
neuron, a decrease of NDD effect (as the level of depression increases)
implies a better response of the neuron to signal detection. Moreover,
since the jitter also decreases for large $\tau_{rec}$, then spike
precision in response to a given external stimulus increases \cite{van2003effects,gutkin2003spike,schneidman_ion_1998}.

\setlength{\intextsep}{10pt}
\begin{figure}[H]
\begin{centering}
\includegraphics[scale=0.25]{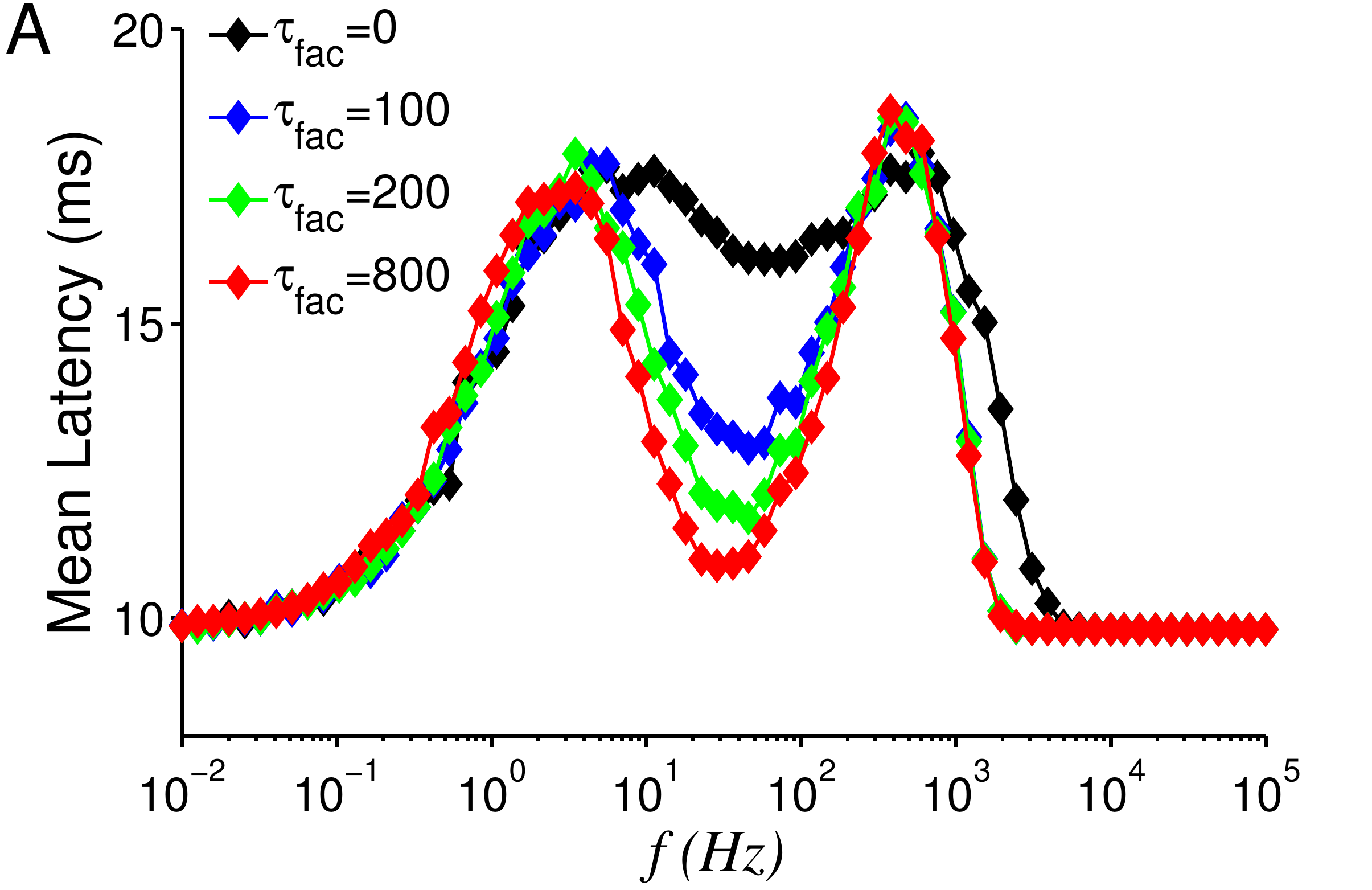}\includegraphics[scale=0.25]{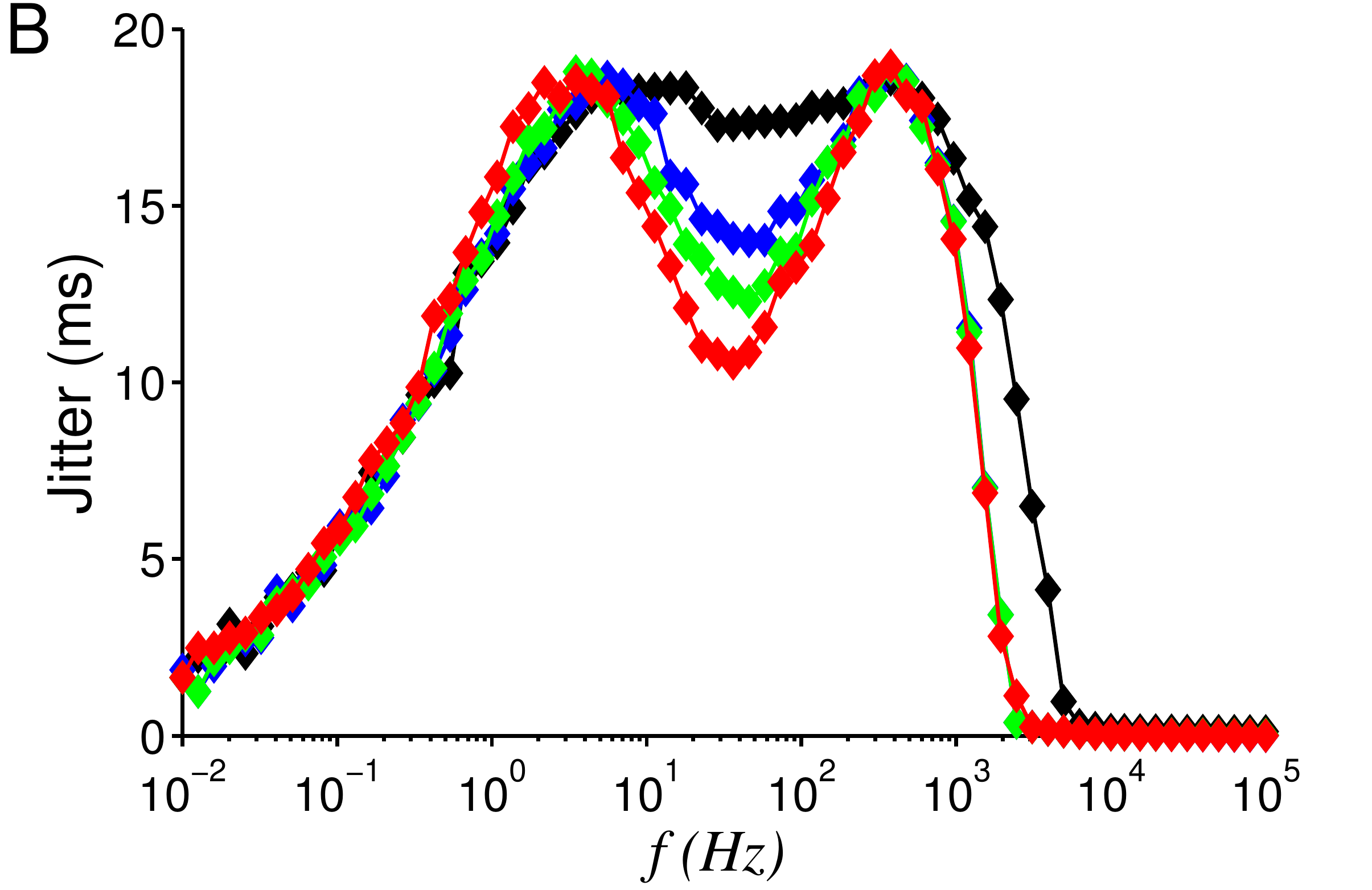}
\par\end{centering}
\centering{}\protect\caption{(Color online) The impact of synaptic facilitation on first-spike
latency statistics. Mean latency and jitter are respectively shown
in panels A and B as a function of $f$ when STF and STD mechanisms
are both present at synapses. As seen in the figure, facilitation
favors the emergence of DNDD behavior. Release probability at rest
and neurotransmitter recovery time constant were set to $\mathcal{U}=0.2$
and $\tau_{rec}=100\, ms$, respectively. Other synaptic parameters
were as in Fig. 2. }
\end{figure}
\setlength{\intextsep}{10pt}

Synapses in the brain can present \textendash{} in addition to synaptic
depression \textendash{} synaptic facilitation mechanism that induces
an enhancement of postsynaptic response at short time scales \cite{tsodyksPNAS,tsodyksNC}.
During synaptic transmission, a complex interplay between these two
mechanisms can occur, a fact that can have strong implications on
first spike latency in response to given stimulus. We investigated
this important issue in our system and the results are depicted in
Fig. 3. This figure shows latency statistics as a function of $f$
for different levels of facilitation, controlled by $\tau_{fac}$,
at a given value of depression $\tau_{rec}=100\, ms.$ The main finding
here is that synaptic facilitation induces a decrease (reaching a
minimum) in mean latency and jitter for a given range of $f$, which
is similar to the case of only depressed synapses with low $\tau_{rec}$,
and that gives rise to emergence of facilitation induced DNDD. Moreover,
such DNDD effect occurs for most values of $\tau_{rec}$ that one
can consider as it is illustrated in Fig. 4. As seen in the figure,
facilitation\\


\setlength{\intextsep}{0pt}
\begin{figure}[H]
\begin{centering}
~~~~~~~\includegraphics[scale=0.25]{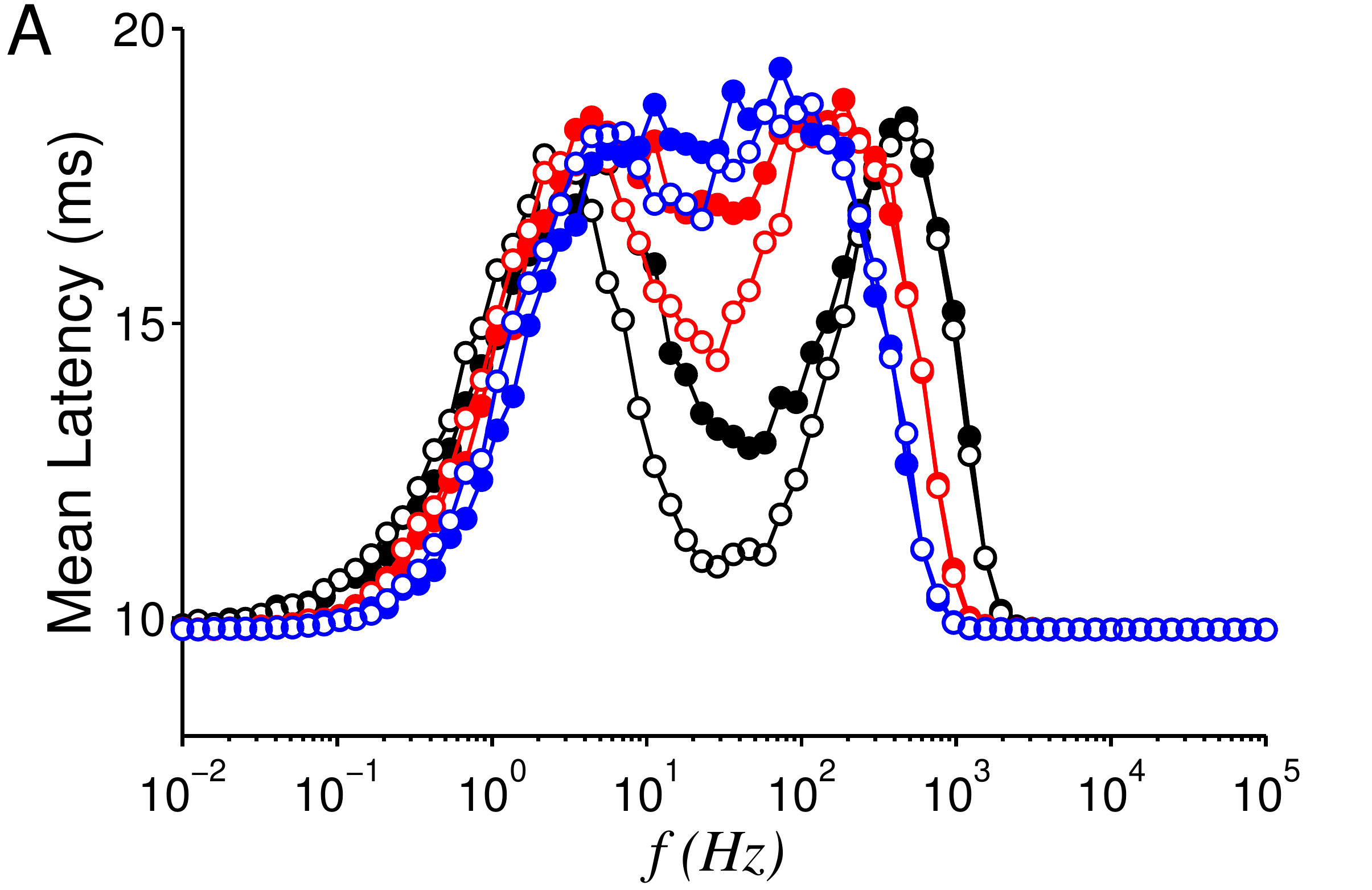}\includegraphics[scale=0.25]{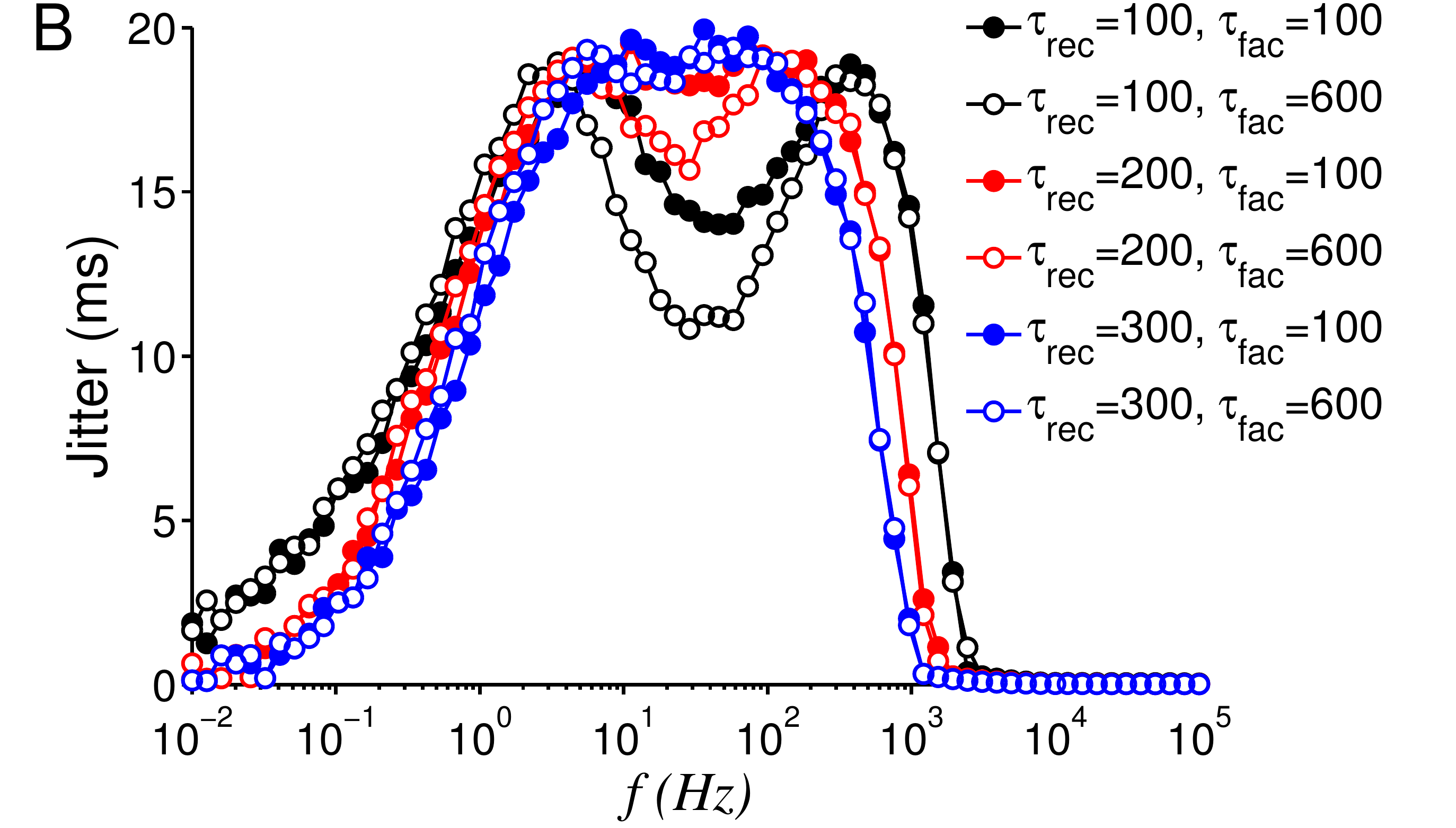}
\par\end{centering}
\protect\caption{(Color online) The influence of competition between STF and STD on
NDD. Panels A and B show, respectively, mean latency and jitter as
a function of $f$ for different sets of $\tau_{rec}$ and $\tau_{fac}$.
For each value of $\tau_{rec}$, we compare the difference in NDD
curves when facilitation time constant is increased from $\tau_{fac}=100\, ms$
to $600\, ms.$ The two mechanisms act in opposite ways: at a given
value of facilitation where DNND effect emerges, an increase of the
level of depression destroys it; on the hand for a given value of
depression, DNDD effect becomes more prominent if facilitation level
of synapses is increased. Other synaptic parameters were as in Fig.
3. }
\end{figure}

\noindent tends to favor the DNDD for the levels of depression
considered in our analysis. However, figure also shows that, for a
given value of facilitation, DNDD effect becomes less evident by enlarging
the level of depression in the system (as in the case of pure depressing
synapses).

On the other hand, assuming $\tau_{in}$ $\ll$ $\tau_{rec}$, $\tau_{fac}$,
it has been shown that release probability at rest, i.e. $\mathcal{\mathcal{U}}$
, is a relevant parameter that can critically control the level of
synaptic depression in the presence of facilitation \cite{tsodyks2005course},
a fact that can induce different intriguing computational implications
\cite{torresNC2007,mejiasCD08,mejias09,Mongillo2012,TorresKappen2013,mejias2011emergence}).
In this context, another important issue concerning our study here
is to investigate the impact of $\mathcal{\mathcal{U}}$ on the emergence
and behavior of NDD as a function of $f$ when synapses present different
types of short-term synaptic plasticity. Our analysis is depicted
in Fig. 5 for depressing and facilitating synapses. When only STD
mechanism is present at synapses with a depression level of $\tau_{rec}=100\, ms$
that induces NDD effect, we have DNDD for large $\mathcal{\mathcal{U}}$,
single NDD for intermediate values, and lack of NDD for very small
values of $\mathcal{\mathcal{U}}$ (see Panels A and B). On the other
hand, for facilitating synapses, we observe similar influence of $\mathcal{\mathcal{U}}$
on NDD behavior, but interestingly, NDD effect is always present regardless
of $\mathcal{\mathcal{U}}$ (see Panels C and D). Moreover, contrary
to the behavior of DNDD curves induced by $\tau_{rec}$ or $\tau_{fac}$
where the second peak is modulated (see Fig. 2 and 3), $\mathcal{\mathcal{U}}$
has more importance on modulating the first peak of DNDD curves. 

\setlength{\intextsep}{10pt}
\begin{figure}[H]
\begin{centering}
\includegraphics[scale=0.25]{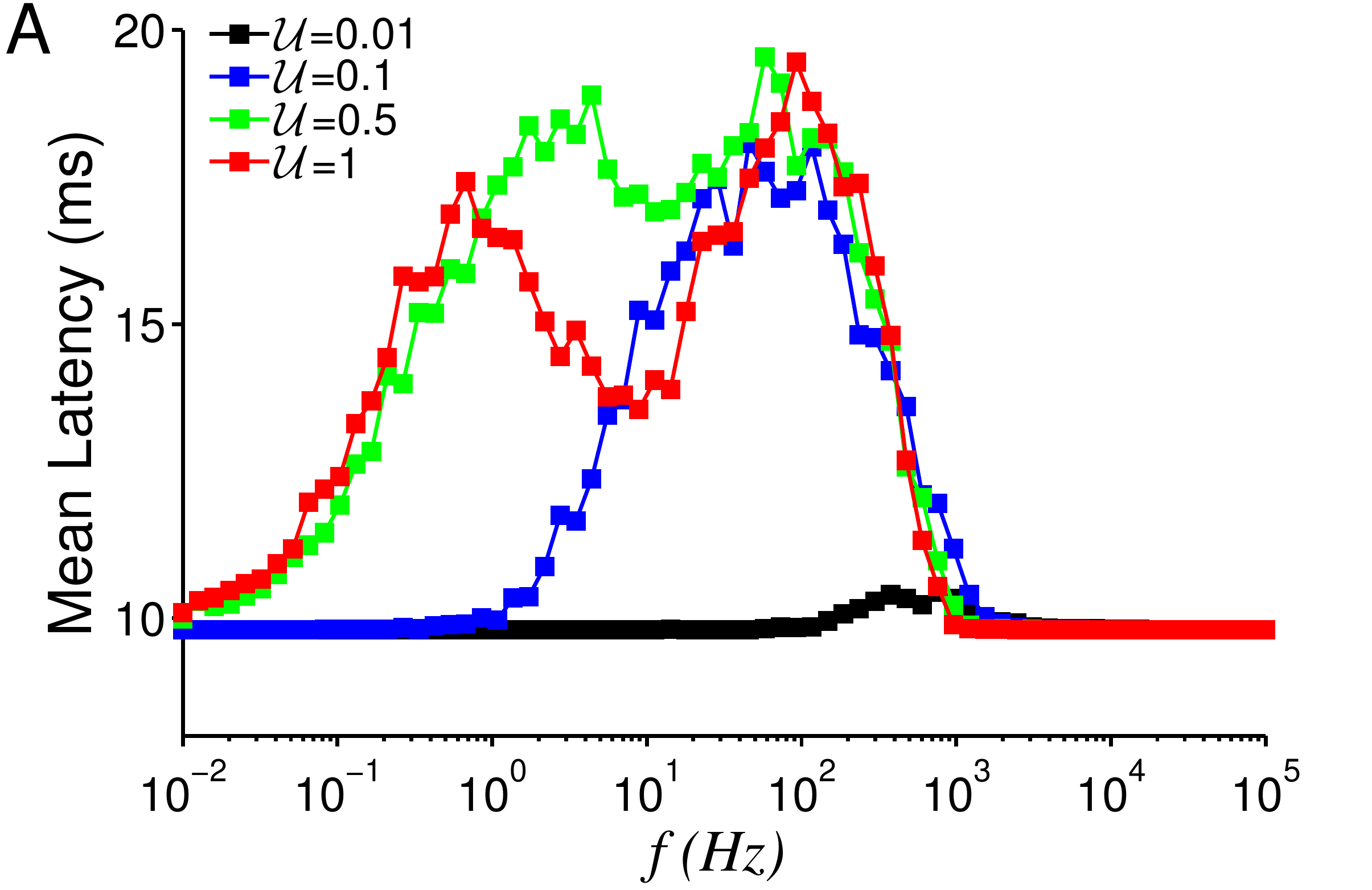}\includegraphics[scale=0.25]{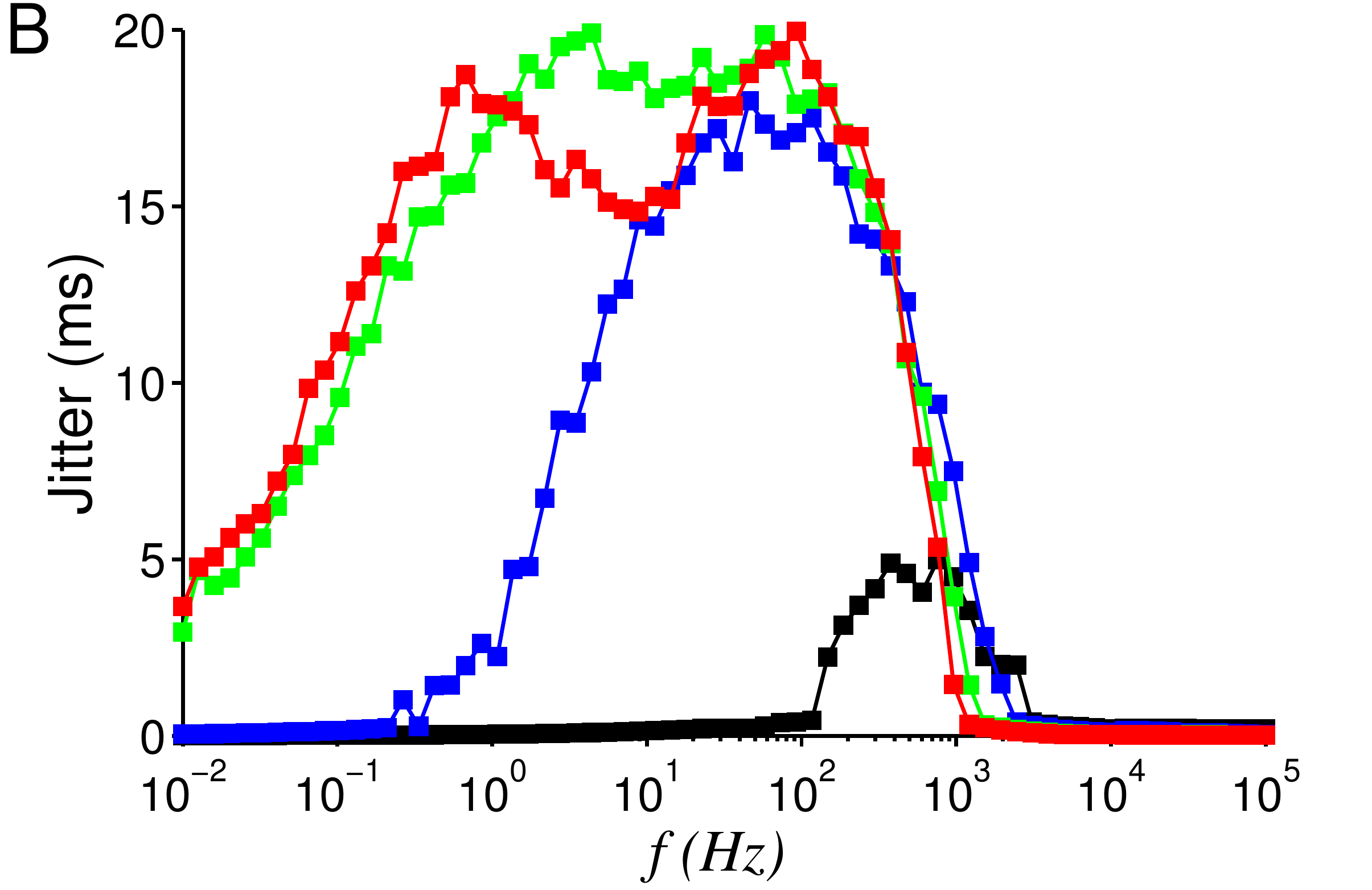}\vspace{1mm}
\par\end{centering}
\begin{centering}
\includegraphics[scale=0.25]{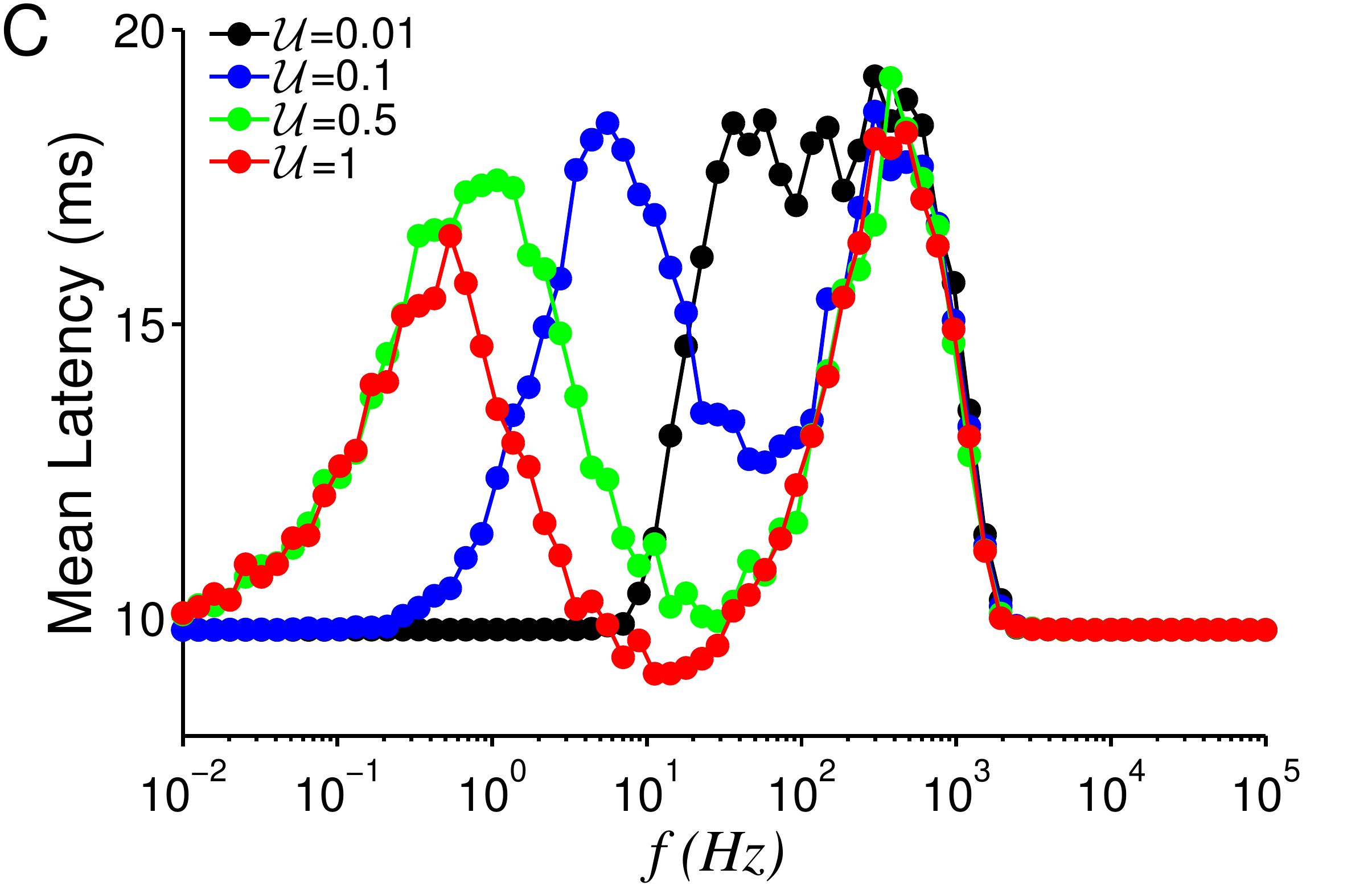}\includegraphics[scale=0.25]{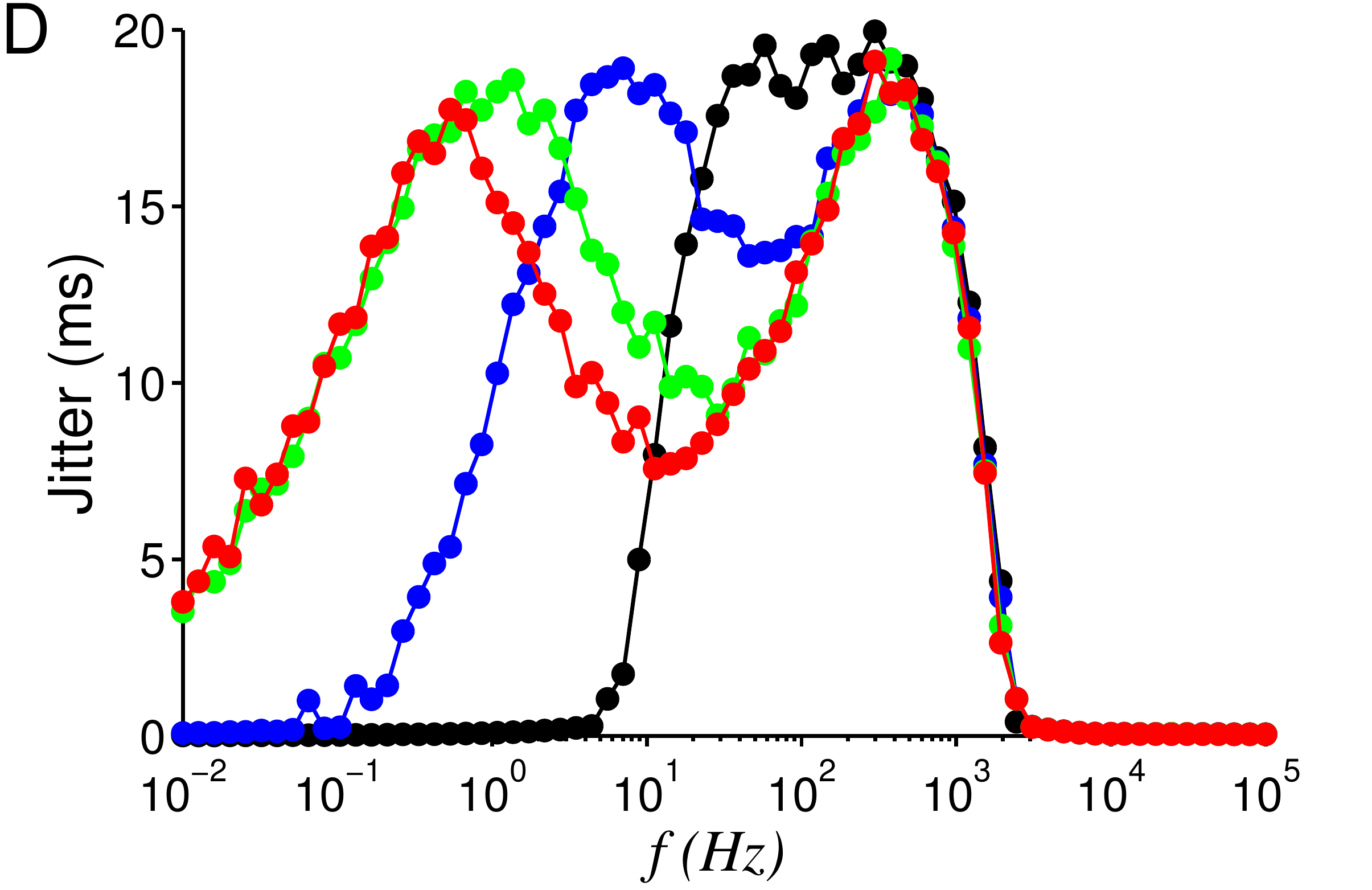}
\par\end{centering}
\protect\caption{(Color online) The influence of release probability at rest $\mathcal{\mathcal{U}}$
on NDD. Panels A and B illustrate, respectively, variations of mean
latency and jitter as a function of $f$ for several different values
of $\mathcal{\mathcal{U}}$ in the presence of only STD mechanism
($\tau_{rec}=100\, ms$ and$\tau_{fac}=0\, ms$). Panels C and D shows
the same statistics when STD and STF mechanisms are both present at
synapses ($\tau_{rec}=100\, ms$ and $\tau_{fac}=300\, ms$). Other
parameters were as in Fig. 2. }
\end{figure}

\subsection{Mechanism of NDD and DNDD}

The underlying mechanism for the emergence of NDD in our system is
the following. First, let us assume that a typical H-H neuron is set
at its resting state in absence of a noisy synaptic input. Let us
consider then that it is subject to a sinusoidal signal which puts
it in a spiking regime such that the neuron generates only an AP within
the positive phase of the signal around its maximum amplitude (i.e.,
one AP per period of the signal). In the presence of a noisy synaptic
input, however, this regime is conserved as far as the appearance
of non-depolarizing noise fluctuations around the maximum signal amplitude
impedes the generation of the AP in the current signal cycle. The
last postpones the generation of AP to the subsequent signal cycles.
That is, neuron can fire then around the maximum amplitude of second
or third positive phase of the signal, skipping first signal cycle
with enough noise. This is clearly shown in Fig. 6. Here the case
of low presynaptic firing rate, e.g. $f=2\, Hz$, corresponds to a
low noise regime at which there is not enough strong non-depolarizing
fluctuations to make postsynaptic neuron to skip toward subsequent
depolarizing phases. In fact, first spikes occur only in the first
cycle for low $f$. When presynaptic firing rate increases, for instance,
to $f=30\, Hz$, synaptic noise intensity also enlarges. In this regime,
firing events of neuron depict some skipping behavior from the first
signal cycle. That is, for a given set of trials, neuron fires in first
cycle predominantly (blue), approximately same amount of times
in second cycle (red), and only a few times in third


\begin{figure}[H]
\begin{centering}
\includegraphics[clip,scale=0.32]{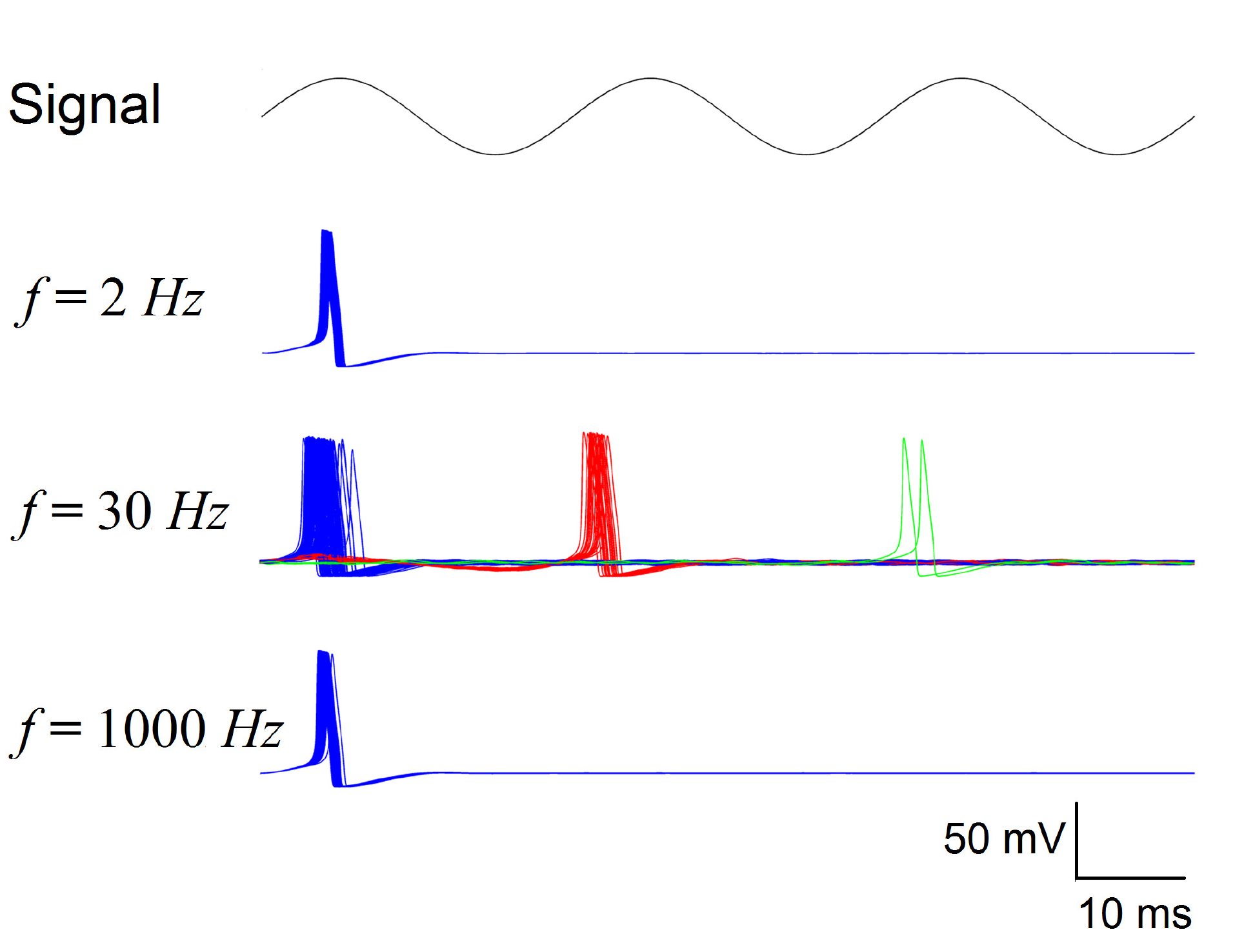}
\par\end{centering}
\protect\caption{(Color online) The mechanism underlying the emergence of NDD. This
figure depicts the position of the first spikes relative to the injected
suprathreshold signal phase for three different values of $f$. In
each panel, we superimpose 50 membrane potential traces obtained from
independent realizations. For each trace, membrane potential is clamped
to the resting state after appearance of first spike. Depending on
the level of $f$, first spike in response to the driving signal might
occur within first (blue), second (red) or even third signal cycle
(green). Neurotransmitter recovery time constant was set to be $\tau_{rec}=100\, ms$
and other synaptic parameters were as in Fig. 2. (Note that vertical
scaling does not refer to the amplitude of periodic signal)}
\end{figure}
\setlength{\intextsep}{0pt}

\noindent cycle (green). This fact induces an increase in mean latency and jitter (see Fig.
2). Finally, for $f=1000\, Hz$, skipping from the first signal cycle
stops and neuron fires again within the first cycle, so mean latency
and jitter start to decrease (see Fig. 2). However, the mechanism
behind this decreasing phase of the NDD curves is different for the
case of depressed ($\tau_{rec}>0$) and non-depressed ($\tau_{rec}=0$)
synapses. For depressed synapses (the case depicted in Fig. 6) and
for large $f$, strength of the synaptic fluctuations decreases which
makes the neuron to fire again in the first signal cycle near the
maximum amplitude, as in the case of very low $f.$ For non-depressed
synapses, however, for all trials neuron fires an AP at the beginning
of first signal cycle due to strong fluctuations (data not shown).
Mean latency and jitter then take small values (see Fig. 2A and B) which
is lower than the noiseless case (very low $f$).

On the other hand, emergence of DNDD behavior in the presence of both
facilitation and depression can be explained with the same skipping
mechanism. This can be seen, for instance, looking at the skipping
behavior of firings depicted in Fig. 7 for $\tau_{fac}=100\, ms$
and $\tau_{rec}=400\, ms$. One can observe that first spikes only
occur within the first signal cycle for very low and very large $f$. 
However, skipping behavior is maximum (which implies larger mean latency and
jitter) for two distinct values of $f$, namely $f=4\, Hz$ and
$500\, Hz$. In fact, for these cases there is a significant amount

\setlength{\intextsep}{10pt}
\begin{figure}[H]
\begin{centering}
\includegraphics[bb=0bp 10bp 555bp 433bp,clip,scale=0.47]{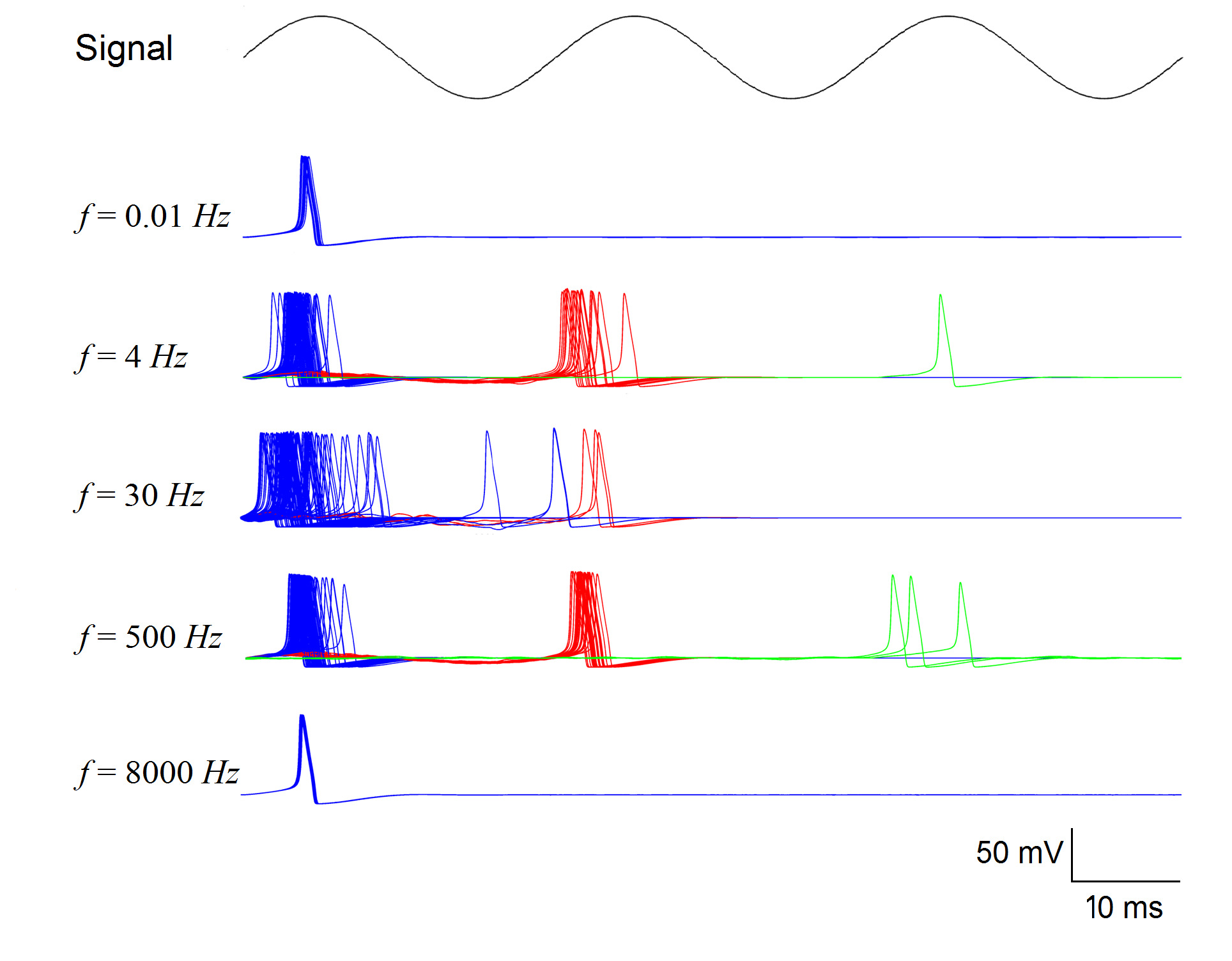}
\par\end{centering}
\protect\caption{(Color online) The mechanism underlying the emergence of DNDD when
facilitation competing with depression is present at synapses. Figure
depicts skipping behavior from the first signal cycle as a function
of $f$. There are two values of $f$ (around $4\, Hz$ and $500\, Hz$)
at which the skipping to second and third signal cycles occurs frequently.
These two points correspond to two maxima of the DNND curve. Synaptic
parameter values were $\tau_{fac}=400\, ms$ and $\tau_{rec}=100\, ms$,
and other parameters were as in Fig. 3. (Note that vertical scaling does
not refer to the amplitude of periodic signal)}
\end{figure}

\noindent of generated spikes during second and third signal cycles (see red
and green voltage traces in Fig. 7). On the other hand, for $f=30\, Hz$
-- which is within these two frequencies that provide maximum skipping
events -- the number of cycle skipped firings starts to decrease resulting
in a minimum of mean latency and jitter among the two previous maxima
(see Fig. 3 and 4). Although, we consider here facilitation induced
DNDD, it is worth noting that the mechanism is the same for only depression
induced DNDD.

\subsection{Non-monotonic frequency dependency of synaptic current fluctuations
shapes NDD and DNDD curves}

Cycle-skipped firings and emergence of NDD in our considered system
are mainly related with the behavior of total synaptic current fluctuations
with $f$ and synapse dynamics. Because of the presence of short-term
plasticity mechanisms at synapses, each individual transmitted AP
in a spike train evokes different postsynaptic current (PSC) responses
depending on the type of plasticity and $f$. As an example in Fig.
8, we illustrate such difference at a single synapse where only STD
mechanism is considered during synaptic transmission. At a certain
level of STD (fixed value of $\tau_{rec}$), it is seen that the amplitude
of generated PSCs in response to each AP does not change very much
for small $f$ because there is enough time between successive APs
to recover neurotransmitter vesicles. However, with increase in $f$,
since presynaptic APs tend to appear more likely within the recovery
time interval ($\tau_{rec}$), vesicles could not be fully recovered
for closely time-spaced successive APs. This indicates a reduction
in PSC amplitudes for those of APs. Or alternatively, for a fixed
$f$, an increase in $\tau_{rec}$ causes more attenuated excitatory
PSCs (EPSC) or inhibitory PSCs (IPSC) due to requirement of more time
for vesicle replacement (see Fig. 8). 

\begin{figure}[H]
\begin{centering}
\includegraphics[scale=0.55]{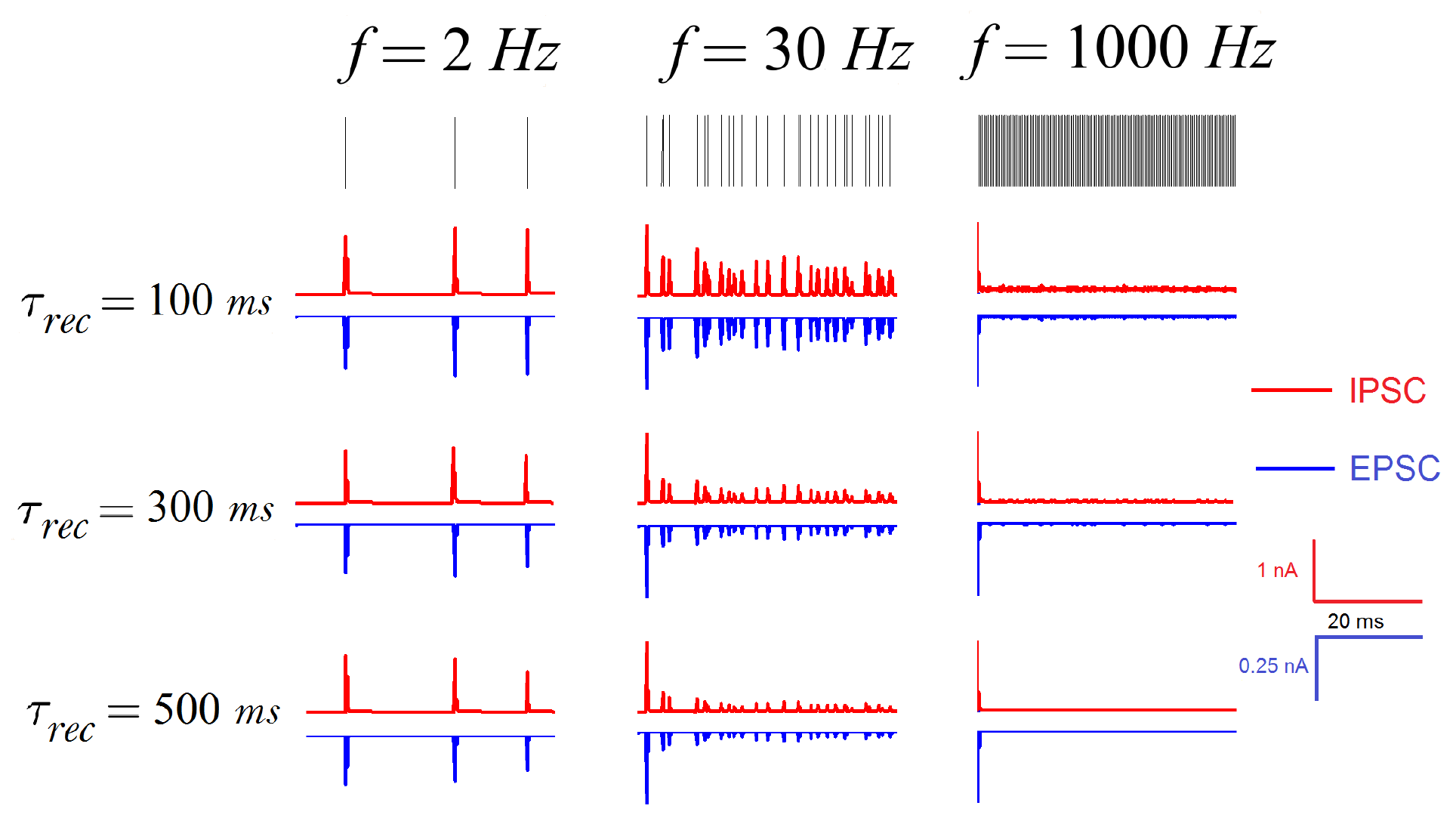}
\par\end{centering}
\protect\caption{(Color online) Postsynaptic currents generated from a presynaptic
Poissonian spike train arriving with different values of $f$ at both
excitatory (red) and inhibitory (blue) synapses that include STD mechanism.
For a fixed depression level, it is clearly seen that PSCs get smaller
as $f$ increases. On the other hand, for a given $f$, an increase
in $\tau_{rec}$ induces attenuation of PSCs.\textbf{ }}
\end{figure}

Since we assume in our study that all synapses present such dynamical
behavior, the resulting total synaptic current introduced into the
postsynaptic neuron will present statistical features that will depend
on both $\tau_{rec}$ and $f$ in a complex way. Moreover, we are
considering a balance between excitation and inhibition, and therefore,
the resulting mean synaptic current $\mu_{I}\approx0$ and will not
change when these parameters are varied. However, the magnitude of
total synaptic current fluctuations, namely $\sigma_{I}$, exhibits
a non-monotonic behavior with $f$ as shown in Fig. 9A for depressing
synapses (see also \cite{mejias2011emergence} and \cite{Romani2006}
for analytical approximation of such dependency), which is more prominent
when $\tau_{rec}$ increases. That is, at low $f$ values, there is
a small number of transmitted spikes that contribute to total synaptic
current. Since these spikes are temporally uncorrelated (they are
Poissonian distributed) and synaptic depression is weak due to low
$f$ (there are not strong differences in the amplitude of the PSC
generated by each arriving spike), it is straightforward to demonstrate
that $\sigma_{I}$ is proportional to $f$ (see Fig. 9A), which then
is small. As $f$ increases up to a moderate value where STD mechanism
still has not a strong effect, $\sigma_{I}$ continues rising monotonically
but slowly because of the appearance of some closely time-spaced spikes
in presynaptic input train which can be then depressed. After this
limiting point of $f$, STD mechanism starts to decrease mean PSCs
seriously that gives rise to a decrease in $\sigma_{I}$ for high
$f$ because of increased number of closely time-spaced spikes. Besides,
increasing the level of depression at synapses $(\tau_{rec}\rightarrow\infty)$
modulates this trend by decreasing the peak value of $\sigma_{I}$
(STD is stronger). 

\begin{figure}
\begin{centering}
\includegraphics[scale=0.25]{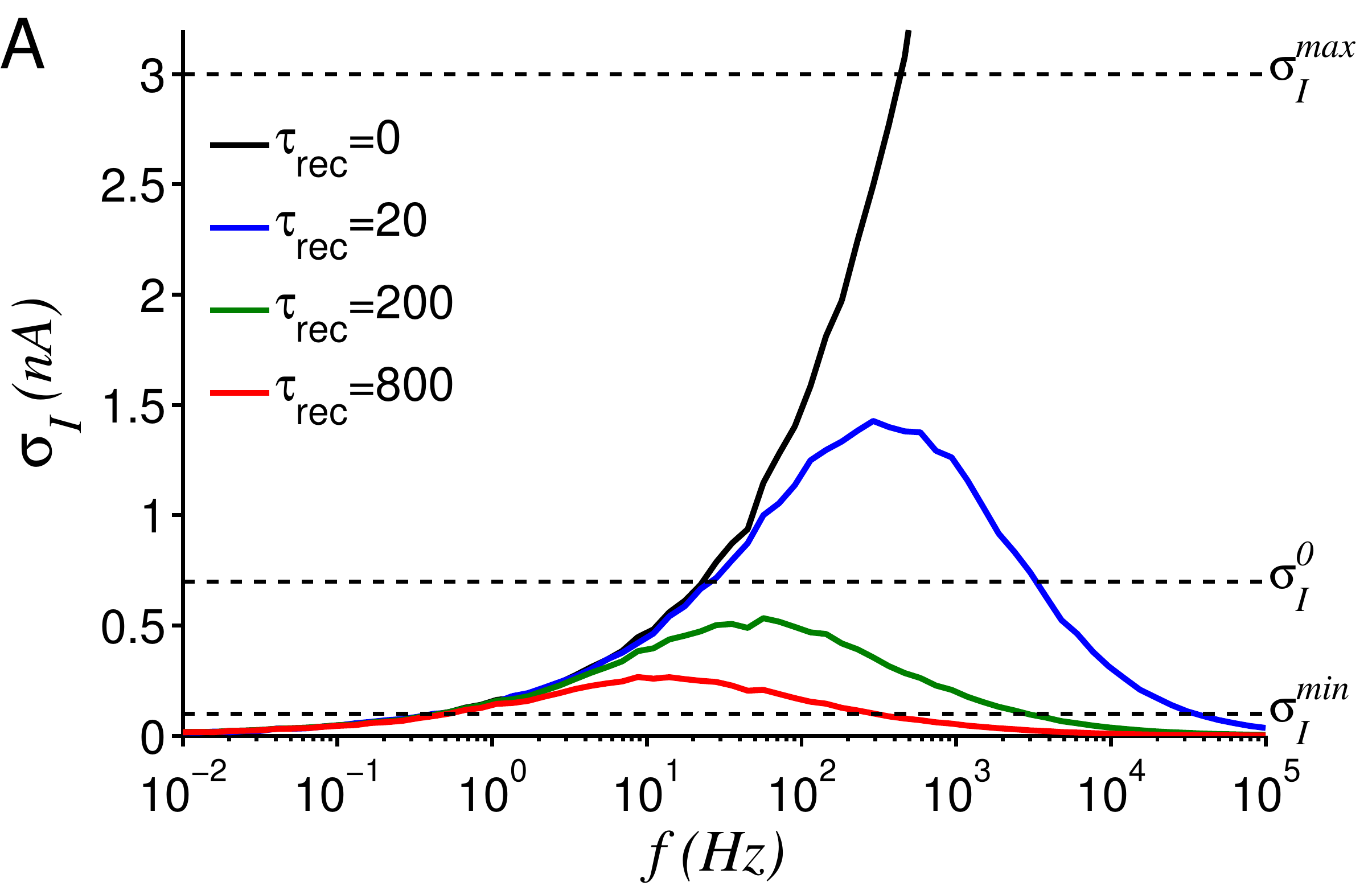}\hspace{2mm}\includegraphics[scale=0.25]{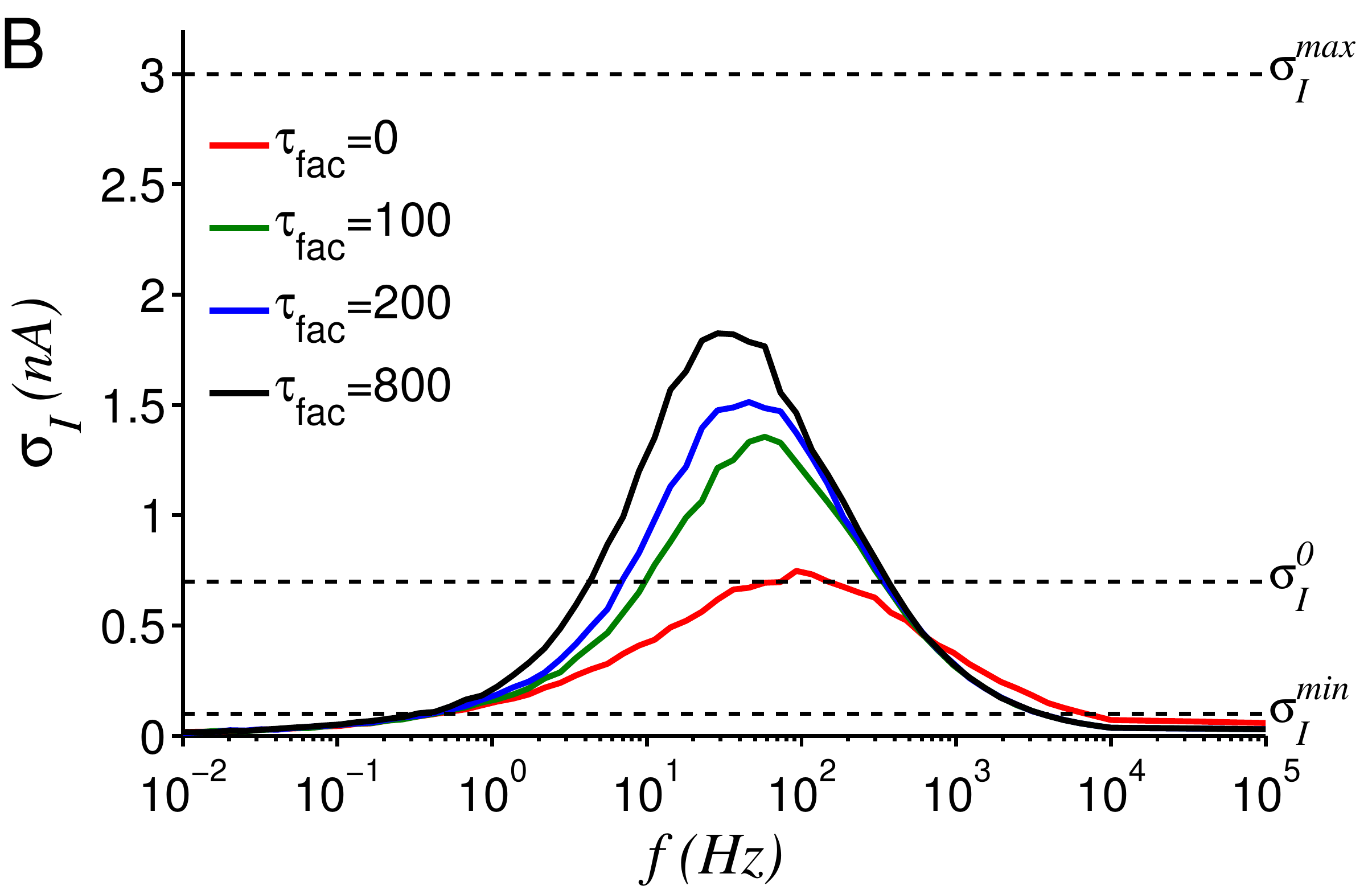}
\par\end{centering}
\protect\caption{(Color online) Variation of synaptic current fluctuations with $f$
when dynamic as well as static synapses are considered. Panel A shows
the dependency of $\sigma_{I}$ on $f$ in the presence of only STD
mechanism at synapses ($\tau_{fac}=0$). Several different values
of $\tau_{rec}$ are considered including $\tau_{rec}=0$ which characterizes
the case of static synapses. In Panel B, variation of $\sigma_{I}$
with $f$ is illustrated when STD and STF mechanism are both present
at synapses. Note that $\tau_{fac}$ is used to control competition
between two mechanism where $\tau_{rec}$ is fixed to $100\, ms$.
The condition $\sigma_{I}=\sigma_{I}^{0}$ marks the emergence of
the two maxima on DNDD curve (see Fig. 3), whereas the maximum of
$\sigma_{I}$ is associated to minimum between these two maxima (see
main text for a more detailed explanation). }
\end{figure}

Behavior of $\sigma_{I}$ with $f$ for different values of $\tau_{rec}$
can explain the shape of NDD curves. When $\tau_{rec}=0$ (static
synapses), since $\sigma_{I}$ rises monotonically with increase in
$f$ (see Fig. 9A), it can fall within an interval $(\sigma_{I}^{min},\,\sigma_{I}^{0})$
for a particular range of $f$, namely ($f_{min},\, f_{0}$), that
induces an increase of the mean latency and jitter based on the mechanism
described above in Fig. 6. Here $\sigma_{I}^{min}$ is by definition
the minimum current fluctuation value that can start first signal
cycle skipping, and $\sigma_{I}^{0}\approx0.7\, nA$ is the value
of current fluctuations at which number of skipping events is maximum
\cite{pankratova2005b}. When $f$ increases further within the interval
$(f_{0},\, f_{max}),$ $\sigma_{I}$ also rises passing through the
interval $(\sigma_{I}^{0},\,\sigma_{I}^{max})$. Here $\sigma_{I}^{max}$
is the level of current fluctuations that stops skipping for large
$f$. Therefore, within the range $(\sigma_{I}^{0},\,\sigma_{I}^{max})$,
$\sigma_{I}$ starts to decrease the number of skipping events (probability
of firings during the first signal cycle tends to increase) and creates
the decay phase of NDD curve. For very large values of $f$ (that
implies $\sigma_{I}\geq\sigma_{I}^{max}$), background activity suppress
the suprathreshold periodic signal, and consequently mean latency
gets lower values than the deterministic case because the response
time is mostly determined by large synaptic current fluctuations at
each trial. 

On the other hand, as STD mechanism takes place at synapses, the optimal
amount of synaptic current fluctuations for cycle-skipped firings
can be achieved for a wide range of $f$ and $\tau_{rec}$ due to
non-monotonic behavior of $\sigma_{I}$ with $f$. More precisely,
as $f$ increases, $\sigma_{I}$ first exceeds the $\sigma_{I}^{min}$
triggering cycle-skipped firings and initiating the rising phase of
the NDD. Then, it starts to decrease after reaching some maximum,
which is determined by the level of depression $\tau_{rec}$ and that
induces the maximum level of NDD curve. Finally, it crosses again
$\sigma_{I}^{min}$ from above which stops cycle-skipped firings and
creates the decay phase of the NDD curve. Therefore, it can be said
that appearance of NDD with depressing synapses depends on $f$ range
that provides the condition of $\sigma_{I}\geq\sigma_{I}^{min}$.
As seen in Fig. 9A, since $f$ range providing this condition starts
to shrink with increase in $\tau_{rec}$, the width and maximum of
NDD curves decrease for more depressed synapses indicating tendency
to disappearance of NDD effect (see Fig. 2A and B). 

The previous scenario described for depressing synapses occurs for
values of $\tau_{rec}$ such that maximum of current fluctuations
is smaller than $\sigma_{I}^{0},$ that happens for $\tau_{rec}\geq100\, ms$
and induces single NDD curves. For values of $\tau_{rec}<100\, ms,$
as we have already mentioned in Section III. A, a DNDD effect emerges.
This can be explained due to the fact that maximum of $\sigma_{I}$
(in the $f$ domain) is larger than $\sigma_{I}^{0}$ for such levels
of depression (see Fig. 9A). This happens since non-monotonic behavior
of $\sigma_{I}$ occurs for very large values of $f$ in such a way
that $\sigma_{I}$ still grows as a function of $f$, for the range
of frequencies at which NDD appears, and may exceed $\sigma_{I}^{0}$.
Then, since $\sigma_{I}^{0}$ corresponds to the strength of the current
fluctuations at which maximum skipping behavior occurs, a larger value
of $\sigma_{I}$ starts to stop skipping events, which creates the
first NDD curve. This takes place until $\sigma_{I}$ reaches its
maximum. When $\sigma_{I}$ decreases from its maximum (for larger
$f$), skipping events start to increase again until $\sigma_{I}$
becomes smaller than $\sigma_{I}^{0}.$ Further increase in $f$ provides
$\sigma_{I}$ to be less than $\sigma_{I}^{0}$ that creates the second
NDD curve due to reduced number of skipping events. The overall behavior
of mean latency and jitter follow thus a DNDD curve as a function
of $f$ (see Fig. 2C and D).

Emergence of DNDD behavior can be also easily understood in terms
of the complex dependence of $\sigma_{I}$ with $f$ , $\tau_{rec}$
and $\tau_{fac}$ when STF is present at synapses competing with STD
(see Fig. 9B). Such dependency profile can also be found in \cite{Mongillo2012}
and \cite{mejias2011emergence} where the behavior of $\sigma_{I}$
qualitatively matches with our Fig. 9B. In fact for the present situation,
$\sigma_{I}$ can be significantly larger than $\sigma_{I}^{0}$ (due
to facilitation) for some range of intermediate values of $f$. This
increase of $\sigma_{I}$ is due the the fact that STF cancels STD
mechanism for such $f$ values -- so $\sigma_{I}$ can be as large
as in the case of static synapses -- and has no effect at higher values
of $f$, so $\sigma_{I}$ decreases due to STD mechanism. In other
words, the rising phase of $\sigma_{I}$, when facilitation is considered,
is mainly determined by STF mechanism. However, the decreasing phase
after a certain $f$ is due to STD mechanism. The non-monotonic dependence
of $\sigma_{I}$ with $f$ makes it to cross the value $\sigma_{I}^{0}$
two times, one with positive and the other with negative slope, both
corresponding then to maxima in mean latency and jitter. The minimum
between these two maxima -- which defines the DNDD curve -- appears
where $\sigma_{I}$ reaches its largest value above $\sigma_{I}^{0}$
since at this value the number of cycle skipping events is smallest
within the frequency range among the two maxima of the DNDD curve.
As larger is the distance above $\sigma_{I}^{0}$ to the maximum of
$\sigma_{I}$ (e.g increasing $\tau_{fac}$), a more deep minimum
can be obtained for the DNDD curve (compare Fig. 3 and Fig. 9B). Moreover,
for very large $\tau_{fac}$ and a given value of $\tau_{rec}$, this
minimum of DNDD saturates and can not be more deep as shown in Fig.
3 because the maximum of $\sigma_{I}$ also saturates for large $\tau_{fac}$.

\subsection{Sensitivity of latency on synaptic current fluctuations}

As a final result of our study, we here present an interesting low-frequency
behavior of mean latency and jitter when depressing synapses are present.
As can be observed by comparing the latency statistics and $\sigma_{I}$
values, respectively, in Fig. 2 and Fig. 9A, although there is no
significant variation in $\sigma_{I}$ for different levels of synaptic
depression at low $f$, mean latency and jitter get different values.
Note that this effect does not emerge when facilitation is present
at synapses (see Fig. 3). To illustrate this fact, we compute the
probability distribution of first spike latencies for three different
values of $\tau_{rec}$ that generate almost equal strengths of $\sigma_{I}$
at $f=2\, Hz$, a value at which difference in latency statistics
is more obvious. As seen in Fig. 10, although amplitudes of synaptic
current fluctuations $\sigma_{I}$ are almost same for different levels
of depression, first spike latency probability distributions are quite
different. For instance, for the lowest level of synaptic depression
considered here, latency probability distribution has a pronounced
second peak around the time for the depolarization phase of second
signal cycle (see inset). Notably, such a multimodal behavior of H-H
model for low noise regime was also reported in \cite{Luccioli06}.
This second peak at low $f$ indicates that a small increase in $\sigma_{I}$
significantly increases the probability for skipping from first cycle
of the stimulus. A higher probability of skipping results in an increased
values of corresponding mean latency and jitter for low $\tau_{rec}$
with respect to other considered synaptic depression levels (where
this second peak in probability distribution is less pronounced or
even absent). 

\begin{figure}[H]
\begin{centering}
\includegraphics[scale=0.3]{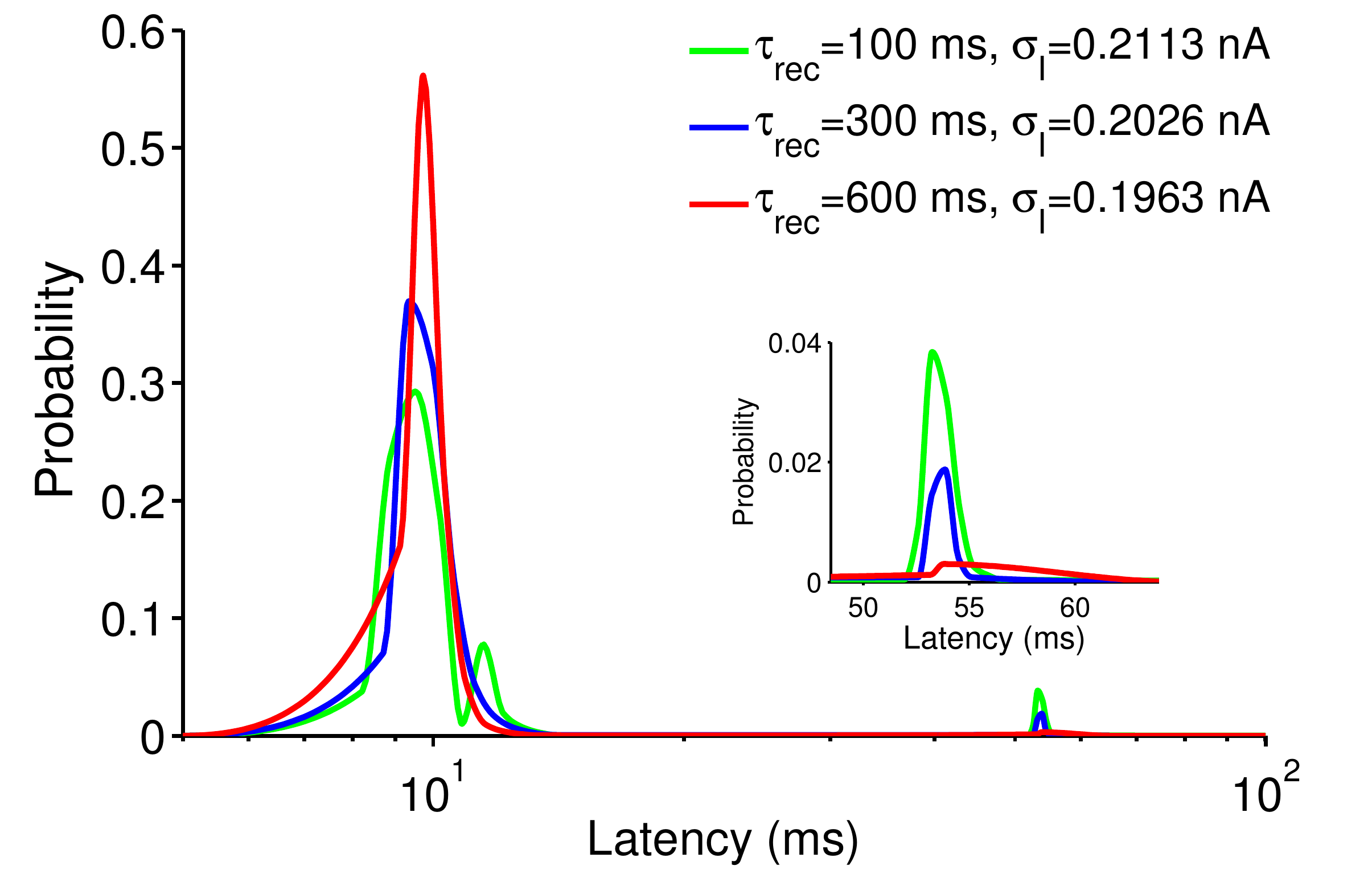}
\par\end{centering}
\protect\caption{(Color online) Probability distribution of first spike latencies for
three different levels of synaptic depression ($\tau_{rec}=100,\,300$
and $600\, ms$) when the presynaptic firing rate was set to $f=2\, Hz.$
Although amplitudes of synaptic current fluctuations $\sigma_{I}$
are almost same, probability distributions of first spike occurrence
time is quite different. Inset shows enlargement of the second peak
that is not clear in the main plot. Other synaptic parameters were
as in Fig. 2.}
\end{figure}

\section{Conclusion}

In this work, we have studied first spike latency response of a H-H
neuron that receives, in addition to a suprathreshold signal, a noisy
current from a finite number of afferents through dynamic synapses.
The noisy input current is originated from arrival of uncorrelated
spikes trains in each afferent that follow a Poissonian statistics
with mean firing frequency $f$. To characterize NDD phenomenon, we
have computed mean latency response and its jitter as a function of
this presynaptic firing rate $f$. This constitutes a more realistic
scenario than traditional studies, since $f$ provides a suitable
biophysically realistic parameter to control the level of activity
in actual neural systems. Our work reveals that when synapses do not
present activity dependent synaptic mechanisms, such as STD nor STF,
classical NDD behavior is found as a function of $f,$ which in this
case (static synapses) is a biologically meaningful measure of noise
level. However, when dynamic synapses with STD and STF are included
in description of synapses, we found a very different behavior of
response latency as a function of $f.$ This is due to the non-monotonic
dependency of current fluctuations on $f$ and how it is modulated
by the dynamic synapse parameters (i.e., $\tau_{rec}$, $\tau_{fac}$
and $\mathcal{\mathcal{U}}$). For instance, for the case of pure
depressing synapses, we found two different behaviors: a new intriguing
DNDD behavior occurs for low levels of STD. This disappears as STD
level increases and single NDD emerges. In addition, we found that
when STD level increases even more in this last case, NDD effect starts
to disappear. This provides possible use of large STD levels as a
mechanism to improve first spike latency coding. However, for intermediate
values of $\tau_{rec}$ ($100-200\, ms$), latency coding efficiency
may be poor because one has large amplitude and width in the frequency
domain of the resulting single NDD curves. On the other hand, in the
presence of STF competing with STD, single NDD or DNDD can emerge
depending on the balance between these two synaptic mechanisms, in
such a way that STF favors the emergence of DNDD whereas STD favors
the existence of single NDD. We have also demonstrated the importance
of release probability at rest $\mathcal{\mathcal{U}}$ on NDD. When
only STD mechanism is present at synapses, it is shown that NDD is
not emerging for very low values of $\mathcal{\mathcal{U}}$, however
both NDD and DNDD can emerge as it is increased. On the other hand,
in the case of STF mechanism, NDD exists in response latency of the
neuron regardless of $\mathcal{\mathcal{U}}$ and it can be transformed
to DNDD as $\mathcal{\mathcal{U}}$ increases. Such effects of $\mathcal{\mathcal{U}}$
are also originated from its influence on shaping total synaptic current
fluctuations as $f$ varies.

An interesting point not considered in the present work concerns the
behavior of NDD in the case of unbalanced synaptic background activity.
Note that in our study, the scaling factor $K$ in Eq. (7) determines
the competition between excitation and inhibition. Variation of this
variable would provide either an excitation or an inhibition dominated
synaptic background activity. For instance, for $K<4$, presynaptic
excitatory neurons would dominate the overall synaptic background
activity resulting in a positive mean synaptic current. In contrast,
for $K>4$, the sign of the mean synaptic current would be negative
due to the large inhibition. For a given set of dynamic synapse model
parameters, mean synaptic current for these two unbalanced situations
quickly saturates to a positive (for excitation dominated) or negative
(for inhibitory dominated) value in $f$ domain (more or less it is
a DC current) due to the nature of dynamic synapses \cite{delarocha05}.
When $K<4$, injection of a such depolarizing positive bias current
would provide a strong stable spiking regime to the postsynaptic neuron
in which influence of synaptic current fluctuations on spike timing
is negligible. Therefore, we expect that NDD effect would be reduced
significantly or completely removed from latency response of the postsynaptic
neuron. On the other hand, in the case of inhibitory dominated background
activity, a negative hyper-polarizing mean synaptic current would
cause a non-spiking regime for the postsynaptic neuron because suprathreshold
driving periodic signal is already very close to the subthreshold
boundary. In such a non-spiking regime where the neuron exhibits subthreshold
oscillations near its resting potential, occurrence of APs would be
determined randomly by fluctuations of synaptic current. Therefore,
one cannot mention about the NDD phenomenon in inhibitory dominated
synaptic input regime due to definition of this phenomenon (resonance
like dependency on noise).

The present study reports for the first time the emergence of DNDD
behavior as a function of a biophysically realistic parameter controlling
the level of activity in a neural medium. DNDD behavior can be
useful for first spike latency coding when one is interested to encode
the stimulus within a particular range of noise in the system. In
fact, a minimum of mean latency among the two maxima might provide
a suitable range of intermediate working frequencies at which the
mean latency is near to the deterministic case. The similar behavior
for the jitter as a function of $f$ (see, e.g., Fig. 3 and 4) indicates
an increase in spike precision that might be useful for some specific
information tasks requiring spike time reliability.

Finally, we would like to note that the results presented in this
work were obtained by considering first, a balance between presynaptic
excitation and inhibition resulting in a zero mean fluctuating total
synaptic current and, second, a critical driving suprathreshold signal
frequency range providing the postsynaptic neuron to operate closely
to its spiking threshold. Under these conditions, due to the mechanism
explained in Section III. B, we expect that any biological process
having capacity to modulate $\sigma_{I}$ non-monotonically between
$\sigma_{I}^{min}$ and $\sigma_{I}^{max}$ might give rise to similar
NDD and DNDD effects on response latency of a postsynaptic neuron.
In this context, a possible extension of our study could be investigating
the influence of astrocytic gliotransmission on the emergence and
behavior of NDD, since it has been shown that activation of presynaptic
receptors by chemicals released by surrounding glia can regulate the
mechanism of short term plasticity in an activity dependent manner
(see \cite{DePitta2015}).  

\begin{acknowledgments}
MU acknowledges financial support from the Scientific and Technological
Research Council of Turkey (T\"{U}B\.{I}TAK) B\.{I}DEB-2219 Postdoctoral Research
program and Bulent Ecevit University research foundation under the
project BAP2013-39971044. JJT acknowledges support from the Spanish
Ministry of Economy and Competitiveness under the project FIS2013-43201-P.
\end{acknowledgments}

\bibliographystyle{unsrt}


\begin{thebibliography}{10}

\bibitem{Eckmiller:90}
R.~Eckmiller, G.~Hartmann, and G.~Hauske, editors.
\newblock {\em Parallel Processing in Neural Systems and Computers}.
\newblock North-Holland, Amsterdam, 1990.

\bibitem{vanrullen_spike_2005}
R.~Vanrullen, R.~Guyonneau, and S.~J. Thorpe.
\newblock Spike times make sense.
\newblock {\em Trends Neurosci.}, 28:4, 2005.

\bibitem{chase2007first}
S.~M. Chase and E.~D. Young.
\newblock First-spike latency information in single neurons increases when
  referenced to population onset.
\newblock {\em Proc. Natl. Acad. Sci. USA}, 104:5175, 2007.

\bibitem{panzeri2001role}
S.~Panzeri, R.~S. Petersen, S.~R. Schultz, M.~Lebedev, and M.~E. Diamond.
\newblock The role of spike timing in the coding of stimulus location in rat
  somatosensory cortex.
\newblock {\em Neuron}, 29:769, 2001.

\bibitem{junek2010olfactory}
S.~Junek, E.~Kludt, F.~Wolf, and D.~Schild.
\newblock Olfactory coding with patterns of response latencies.
\newblock {\em Neuron}, 67:872, 2010.

\bibitem{furukawa2002cortical}
S.~Furukawa and J.~C. Middlebrooks.
\newblock Cortical representation of auditory space: information-bearing
  features of spike patterns.
\newblock {\em J. Neurophysiol.}, 87:1749, 2002.

\bibitem{heil2004first}
P.~Heil.
\newblock First-spike latency of auditory neurons revisited.
\newblock {\em Curr. Opin. Neurobiol.}, 14:461, 2004.

\bibitem{gawne1996latency}
T.~J. Gawne, T.~W. Kjaer, and B.~J. Richmond.
\newblock Latency: another potential code for feature binding in striate
  cortex.
\newblock {\em J. Neurophysiol.}, 76:1356, 1996.

\bibitem{reich2001temporal}
D.~S. Reich, F.~Mechler, and J.~D. Victor.
\newblock Temporal coding of contrast in primary visual cortex: when, what, and
  why.
\newblock {\em J. Neurophysiol.}, 85:1039, 2001.

\bibitem{pankratova2005a}
E.~V. Pankratova, A.~V. Polovinkin, and B.~Spagnolo.
\newblock Suppression of noise in fitzhugh-nagumo model driven by a strong
  periodic signal.
\newblock {\em Phys. Lett. A}, 344:43, 2005.

\bibitem{pankratova2005b}
E.~V. Pankratova and A.~V. Polovinkin.
\newblock {Resonant activation in a stochastic Hodgkin-Huxley model: interplay
  between noise and suprathreshold driving effects}.
\newblock {\em Eur. Phys. J. B}, 45:391, 2005.

\bibitem{ozer2008}
M.~Ozer and L.~J. Graham.
\newblock {Impact of network activity on noise delayed spiking for a
  Hodgkin-Huxley model}.
\newblock {\em Eur. Phys. J. B}, 61:499, 2008.

\bibitem{Ozer2008b}
M.~Ozer and M.~Uzuntarla.
\newblock Effects of the network structure and coupling strength on the
  noise-induced response delay of a neuronal network.
\newblock {\em Phys. Let. A}, 372:4603, 2008.

\bibitem{ozer2009}
M.~Ozer, M.~Uzuntarla, M.~Perc, and L.~J. Graham.
\newblock {Spike latency and jitter of neuronal membrane patches with
  stochastic Hodgkin-Huxley channels}.
\newblock {\em J. Theor. Biol.}, 261:83, 2009.

\bibitem{uzuntarla2012}
M.~Uzuntarla, M.~Ozer, and D.~Q. Guo.
\newblock Controlling the first-spike latency response of a single neuron via
  unreliable synaptic transmission.
\newblock {\em Eur. Phys. J. B}, 85:282, 2012.

\bibitem{fssjn03}
K.~M. Franks, C.~F. Stevens, and T.~J. Sejnowski.
\newblock Independent sources of quantal variability at single glutamatergic
  synapses.
\newblock {\em J. Neurosci.}, 23:3186, 2003.

\bibitem{tsodyksCODING}
G.~Fuhrmann, I.~Segev, H.~Markram, and M.~Tsodyks.
\newblock Coding of temporal information by activity-dependent synapses.
\newblock {\em J. Neurophysiol.}, 87:140, 2001.

\bibitem{markramPNAS}
H.~Markram, Y.~Wang, and M.~Tsodyks.
\newblock Differential signaling via the same axon of neocortical pyramidal
  neurons.
\newblock {\em Proc. Natl. Acad. Sci. USA}, 95:5323, 1998.

\bibitem{markram06}
Y.~Wang, H.~Markram, P.~H. Goodman, T.~K. Berger, J.~Ma, and P.~S.
  Goldman-Rakic.
\newblock Heterogeneity in the pyramidal network of the medial prefrontal
  cortex.
\newblock {\em Nat. Neurosci.}, 9:534, 2006.

\bibitem{bertramJNEURO}
R.~Bertram, A.~Sherman, and E.~F. Stanley.
\newblock Single-domain/bound calcium hypothesis of transmitter release and
  facilitation.
\newblock {\em J. Neurophysiol.}, 75:1919, 1996.

\bibitem{abott1997}
L.~F. Abbott, J.~A. Varela, K.~Sen, and S.~B. Nelson.
\newblock Synaptic depression and cortical gain control.
\newblock {\em Science}, 275:220, 1997.

\bibitem{bibitchkov2002}
D.~Bibitchkov, J.~M. Herrmann, and T.~Geisel.
\newblock Pattern storage and processing in attractor networks with short-time
  synaptic dynamics.
\newblock {\em Network: Comp. Neural}, 13:115, 2002.

\bibitem{torres2002}
J.~J. Torres, L.~Pantic, and H.~J. Kappen.
\newblock Storage capacity of attractor neural networks with depressing
  synapses.
\newblock {\em Phys. Rev. E.}, 66:061910, 2002.

\bibitem{mejias09}
J.~F. Mejias and J.~J. Torres.
\newblock Maximum memory capacity on neural networks with short-term depression
  and facilitation.
\newblock {\em Neural Comput.}, 21:851, 2009.

\bibitem{pantic2003}
L.~Pantic, J.~J. Torres, and H.~J. Kappen.
\newblock Coincidence detection with dynamic synapses.
\newblock {\em Network: Comp. Neural}, 14:17, 2003.

\bibitem{mejiasCD08}
J.~F. Mejias and J.~J. Torres.
\newblock The role of synaptic facilitation in spike coincidence detection.
\newblock {\em J. Comput. Neurosci.}, 24:222, 2008.

\bibitem{Sejnowski2006}
J.~Mishra, J.~M. Fellous, and T.~J. Sejnowski.
\newblock Selective attention through phase relationship of excitatory and
  inhibitory input synchrony in a model cortical neuron.
\newblock {\em Neural Networks}, 19:1329, 2006.

\bibitem{BT05}
C.~I. Buia and P.~H.~E. Tiesinga.
\newblock Rapid temporal modulation of synchrony in cortical interneuron
  networks with synaptic plasticity.
\newblock {\em Neurocomputing}, 65:809, 2005.

\bibitem{TorresKappen2013}
J.~J. Torres and J.~H. Kappen.
\newblock Emerging phenomena in neural networks with dynamic synapses and their
  computational implications.
\newblock {\em Front. Comput. Neurosci.}, 7:30, 2013.

\bibitem{HHb}
A.~L. Hodgkin and A.~F. Huxley.
\newblock A quantitative description of membrane current and its application to
  conduction and excitation in nerve.
\newblock {\em J. Physiol.}, 117:500, 1952.

\bibitem{Braitenberg91}
V.~Braitenberg and A.~Schuz.
\newblock {\em Anatomy of the cortex: statistics and geometry}.
\newblock Springer, Berlin, 1991.

\bibitem{tsodyksNC}
M.~V. Tsodyks, K.~Pawelzik, and H.~Markram.
\newblock Neural networks with dynamic synapses.
\newblock {\em Neural Comput.}, 10:821, 1998.

\bibitem{Fitzpatrick2001}
J.~S. Fitzpatrick, G.~Akopian, and J.~P. Walsh.
\newblock Short-term plasticity at inhibitory synapses in rat striatum and its
  effects on striatal output.
\newblock {\em J. Neurophysiol.}, 85:2088, 2001.

\bibitem{Tecuapetla2007}
F.~Tecuapetla, L.~Carrillo-Reid, J.~Bargas, and E.~Galarraga.
\newblock Dopaminergic modulation of short-term synaptic plasticity at striatal
  inhibitory synapses.
\newblock {\em Proc. Natl. Acad. Sci. USA}, 104:10258, 2007.

\bibitem{Ma2012}
Y.~Ma, H.~Hu, and A.~Agmon.
\newblock Short-term plasticity of unitary inhibitory-to-inhibitory synapses
  depends on the presynaptic interneuron subtype.
\newblock {\em J. Neurosci.}, 32:983, 2012.

\bibitem{Flores2015}
J.~B. Flores, M.~A.~H. Valdez, V.~G.~L. Huerta, E.~Galarraga, and J.~Bargas.
\newblock Diverse short-term dynamics of inhibitory synapses converging on
  striatal projection neurons: differential changes in a rodent model of
  parkinson's disease.
\newblock {\em Neural Plast.}, 2015:573543, 2015.

\bibitem{tsodyksjn00}
M.~Tsodyks, A.~Uziel, and H.~Markram.
\newblock Synchrony generation in recurrent networks with frequency-dependent
  synapses.
\newblock {\em J. Neurosci.}, 20:RC50, 2000.

\bibitem{tsodyksPNAS}
M.~V. Tsodyks and H.~Markram.
\newblock The neural code between neocortical pyramidal neurons depends on
  neurotransmitter release probability.
\newblock {\em Proc. Natl. Acad. Sci. USA}, 94:719, 1997.

\bibitem{van2003effects}
M.~C.~W. Van~Rossum, B.~J. O'Brien, and R.~G. Smith.
\newblock Effects of noise on the spike timing precision of retinal ganglion
  cells.
\newblock {\em J. Neurophysiol.}, 89:2406, 2003.

\bibitem{gutkin2003spike}
B.~Gutkin, G.~B. Ermentrout, and M.~Rudolph.
\newblock Spike generating dynamics and the conditions for spike-time precision
  in cortical neurons.
\newblock {\em J. Comput. Neurosci.}, 15:91, 2003.

\bibitem{schneidman_ion_1998}
E.~Schneidman, B.~Freedman, and I.~Segev.
\newblock Ion channel stochasticity may be critical in determining the
  reliability and precision of spike timing.
\newblock {\em Neural Comput.}, 10:1679, 1998.

\bibitem{tsodyks2005course}
M.~Tsodyks.
\newblock Course 7 activity-dependent transmission in neocortical synapses.
\newblock {\em Les Houches}, 80:245, 2005.

\bibitem{torresNC2007}
J.~J. Torres, J.~M. Cortes, J.~Marro, and H.~J. Kappen.
\newblock Competition between synaptic depression and facilitation in attractor
  neural networks.
\newblock {\em Neural Comput.}, 19:2739, 2008.

\bibitem{Mongillo2012}
G.~Mongillo, D.~Hansel, and C.~van Vreeswijk.
\newblock Bistability and spatiotemporal irregularity in neuronal networks with
  nonlinear synaptic transmission.
\newblock {\em Phys. Rev. Lett.}, 108:158101, 2012.

\bibitem{mejias2011emergence}
J.~F. Mejias and J.~J. Torres.
\newblock Emergence of resonances in neural systems: the interplay between
  adaptive threshold and short-term synaptic plasticity.
\newblock {\em PLoS ONE}, 6:e17255, 2011.

\bibitem{Romani2006}
S.~Romani, D.~J. Amit, and G.~Mongillo.
\newblock Mean-field analysis of selective persistent activity in presence of
  short-term synaptic depression.
\newblock {\em J. Comput. Neurosci.}, 20:201, 2006.

\bibitem{Luccioli06}
S.~Luccioli, T.~Kreuz, and A.~Torcini.
\newblock {Dynamical response of the Hodgkin-Huxley model in the high-input
  regime}.
\newblock {\em Phys. Rev. E}, 73:041902, 2006.

\bibitem{delarocha05}
J.~de~la Rocha and N.~Parga.
\newblock Short-term synaptic depression causes a non-monotonic response to
  correlated stimuli.
\newblock {\em J. Neurosci.}, 25:8416, 2005.

\bibitem{DePitta2015}
M.~De Pitt\^a, N.~Brunel, and A.~Volterra.
\newblock Astrocytes: Orchestrating synaptic plasticity?
\newblock {\em Neuroscience, In press}, 2015.

\end{thebibliography}

\end{document}